\documentclass[12pt]{article}
\usepackage{amssymb,amsmath,epsfig,cite,appendix}
 \allowdisplaybreaks

\begin{document}
\title{\bf  Non-Singular Bouncing Model in Energy Momentum Squared Gravity}

\author{Z. Yousaf$^1$ \thanks{zeeshan.math@pu.edu.pk}, M. Z. Bhatti$^1$ \thanks{mzaeem.math@pu.edu.pk}, H. Aman$^1$
\thanks{huzaifaaman971@gmail.com}, P.K. Sahoo$^2$ \thanks{pksahoo@hyderabad.bits-pilani.ac.in }\\
$^1$Department of Mathematics, University of the Punjab,\\
Quaid-i-Azam Campus, Lahore-54590, Pakistan\\
$^2$Department of Mathematics, \\ Birla Institute of Technology and
Science-Pilani,\\ Hyderabad Campus, Hyderabad-500078, India.}

\date{}
\maketitle
\begin{abstract}
This work is concerned to study the bouncing nature of the universe
for an isotropic configuration of fluid $\mathcal{T}_{\alpha\beta}$
and Friedmann-Lema\^{i}tre-Robertson-Walker metric scheme. This work
is carried out under the novel $f(\mathcal{G},\mathcal{T}_{\alpha
\beta} \mathcal{T}^{\alpha \beta})$ gravitation by assuming a
specific model i.e, $f(\mathcal{G},\mathcal{T}^2)=\mathcal{G}+\alpha
\mathcal{G}^2+2\lambda \mathcal{T}^2$ with $\alpha$ and $\lambda$
are constants, serving as free parameters. {The terms
$\mathcal{G}$ and $\mathcal{T}^2$ served as an Gauss-Bonnet
invariant and square of the energy-momentum trace term as an
inclusion in the gravitational action respectively, and is
proportional to $\mathcal{T}^2=\mathcal{T}_{\alpha \beta}
\mathcal{T}^{\alpha \beta}$.} A specific functional form of the Hubble
parameter is taken to provide the evolution of cosmographic
parameters. A well known equation of state
 parameter, $\omega(t)=-\frac{k \log (t+\epsilon
)}{t}-1$ is used to represent the dynamical behavior of energy
density, matter pressure and energy conditions. A detailed graphical
analysis is also provided to review the bounce. Furthermore, all
free parameters are set in a way, to make the supposed Hubble
parameter act as the bouncing solution and ensure the viability of
energy conditions. Conclusively, all necessary conditions for a
bouncing model are checked.
\end{abstract}
{\bf Keywords:} cosmography; Hubble Parameter;
$f(\mathcal{G},\mathcal{T}_{\alpha \beta} \mathcal{T}^{\alpha
\beta})$. 
\section{Introduction}
According to the big bang hypothesis, the whole universe was created
by a single explosion, with all matter in the cosmos as an
infinite  speck \cite{hogan1998little,guth2001eternal}. This
hypothesis works well in order to study the beginning, but lack to
define different cosmological problems. These problems include the
horizon problem, the flatness problem, the singularity problem, etc.
In order to resolve these big cosmic challenges, different cosmic
theories have been developed in literature
\cite{padmanabhan1988does,earman1999critical,ijjas2018bouncing}.
The bouncing hypothesis is one of the major independent theories that
came up with the answers related to the starting of the universe and
should be enough to resolve the major cosmic problem of singularity.
The bouncing cosmology works on the scheme of an oscillatory universe,
i.e, a universe that came into being from the pre-existing universe
without undergoing the singularity
\cite{alesci2017cosmological,das2018cosmological,mielczarek2010observational}.
This whole transition of the universe not only explains the big-bang
cosmology but also reduces one of the major issues. For the bouncing,
the universe moves into the contraction phase as a matter-dominated
the era of the universe. After the contraction, the universe starts to
expand in a nonsingular manner for which gravity dominates the
matter \cite{cai2013anisotropy,cai2009evolution}. Also, density
perturbations can be produced during the bounce era. This idea of
the origination of the universe is highly accepted and appreciated in
literature.

General relativity ($\mathcal{GR}$) was presented by Einstein and it
was thought to be one of the best theories to explain different
cosmological issues. It explains the gravity under the fabric of
space-time. However, to understand gravity much more effectively and
to provide the answers to the effect of gravity, dark energy, and
accelerated  expansion of the universe under the addition of
different scalar fields, different attempts have been made in past
to modify $\mathcal{GR}$. These modifications change the geometric
or matter or both  parts of the Einstein field equations
accordingly. These could help to discuss the effects of couplings of
matter and curvature terms on the above-described items.
Roshan and Shojai \cite{roshan2016energy} presented the
nonlinear form of matter term i.e, $\mathcal{T}^2 =
\mathcal{T}_{\alpha \beta} \mathcal{T}^{\alpha \beta}$, naming it
$f(\mathcal{R},\mathcal{T^2})$. They further indicated that the use
of nonlinear terms may provide the prevention of early time
singularities. Since the functional form of curvature terms has
helped to introduce new gravitational theories, so it was considered
to be effective to modify the generic action integral of
$\mathcal{GR}$ as corrections. These modifications give light to the
$f(\mathcal{G})$ theory, for which the term $\mathcal{G}$ is defined
as $\mathcal{G} =
\mathcal{R}_{\xi\zeta\alpha\beta}\mathcal{R}^{\xi\zeta\alpha\beta}
-4\mathcal{R}_{\xi\zeta}\mathcal{R}^{\xi\zeta}+\mathcal{R}^2$.
Nojiri and Odintsov \cite{nojiri2005modified} introduced this
$f(\mathcal{G})$ theory for the first time in their work. They
tested solar systems for this formalism and reported the phase
change of acceleration to deceleration for the achievement of
phantom line, which cooperated to study dark energy. Odintsov and
Oikonomou \cite{astashenok2015modified} considered
$\mathcal{R}+f(\mathcal{G})$ form of the gravitational theory to
provide their contribution to the study of gravitational
baryogenesis. Their work included the higher-order derivatives of
Gauss-Bonnet terms that work in order to produce the baryon
asymmetry. Sharif and Ikram \cite{sharif2016energy} gives rise to a
new theory by following the footsteps of Harko. They coupled the
matter part $\mathcal{T}$ with the geometric part of the
$f(\mathcal{G})$ theory, making it $f(\mathcal{G},\mathcal{T})$
cosmology. They investigated the validity of their theory with the
help of energy conditions. Later on, Bhatti \emph{et
al}.\cite{bhatti2021electromagnetic} worked on the
$f(\mathcal{G},\mathcal{T})$ theory to carry out the investigation
of some physically feasible features of compact star formation. They
inferred that the compactness of a star model grows at the
core whereas the energy conditions remain constant. Yousaf and his
mates \cite{yousaf2022f} inspired by \cite{katirci2014f}, have
recently developed a novel $f(\mathcal{G},\mathcal{T}^2)$ to present
the complexity of structural scalars from the use of Herrera's
method of splitting scalars. They considered the exponential
coupling of Gauss-Bonnet terms as a functional form as
$f(\mathcal{G},\mathcal{T}^2)=\alpha \mathcal{G}^n (\beta
\mathcal{G}^m + 1)+\eta \mathcal{T}^2$, to explore the validity of
their solutions for the Darmois and Israel conditions. They also
worked on the non-static complex structures under the same theory to
describe the effects of an electromagnetic field. They used specific
model configuration i.e, $f(\mathcal{G},\mathcal{T}^2)= k_1
\mathcal{G}^{m}(k_2 \mathcal{G}^{n} + 1) + \lambda \mathcal{T}^2$,
in their work. 

Bouncing cosmology has gained much reputation over the past few
years, because of its independent hypothetical nature from different
standard comic problems. Guth \cite{guth2007eternal} during
$1980's$, had put forward his inflationary theory to tackle early
and late time cosmic evolutionary problems. He remained successful
in solving some related problems, but the answer to the initial
singularity is still under concern. One of the best hypotheses to
answer the singularity problem is the bouncing nature of the
universe. The nature of the bouncing universe allows a certain
universe model to transit from a pre-big crunch (contracted) phase
into a new big bang (expanded) phase with the exclusion of
singularity during the whole event \cite{steinhardt2002cosmic}.
Steinhardt and Ijjas \cite{ijjas2017fully} are considered to be the
pioneers of the bouncing hypothesis. They devised a wedge diagram
for a smooth bouncing method to explore the consequences of some
cosmological problems. Sahoo \emph{et al}.
\cite{bhattacharjee2020comprehensive} worked on the non-singular
bouncing by assuming the specific coupling of $\mathcal{R}$ and
$\mathcal{T}$ as $f(\mathcal{R},\mathcal{T})=\mathcal{R} + \chi
\mathcal{R} \mathcal{T}$, for $0<\chi<\frac{\pi}{4}$. They allowed
such a parametric approach for the Hubble parameter to provide no
singularity during the bounce. They used quintom and phantom scalar
field configurations for the bouncing paradigm. Bamba and his
collaborators \cite{bamba2014bouncing} inspected the
singularity-free concept of bounce by considering an exponential
form of scale factor $a(t)=\sigma \exp(\lambda t) + \tau
\exp(\lambda t)$ under the effect of $f(\mathcal{G})$ gravity. They
checked the stability of their assumed solution under the restricted
parametric scheme. 

Yousaf \emph{et al}.
\cite{yousaf2022cosmic,yousaf2022bouncing} explored the bouncing
universe with a specific functional form of Hubble parameter by
taking exponential $f(\mathcal{G},\mathcal{T})$ form. Different
cosmic models are under consideration for the scale factor in order
to determine the value of expansion and contraction at the current
cosmic phase and also to predict the current phase equation of
state. These models predicted different results in the literature.
However, cosmography provided us a benefit in processing
cosmological data for explaining the universal kinematics without
the involvement of the gravity model and hence provided that the
cosmography can be employed with the Taylor expansions as an
alternative scheme. Also, the cosmographic analysis for the
$\mathcal{FLRW}$ universe, is helpful in such a way that it can put
aside the effect of the dynamical field equations
\cite{visser2004jerk}. Gruber \emph{et al}.
\cite{gruber2014cosmographic} studied an alternative approach to
describe cosmography by extending the conventional methodology. They
resulted from numerical values of the cosmographic parameters by
applying the $Pad\acute{e}$ approximations. The testing of the
$\Lambda$CDM model had been conveyed by Busti \emph{et al}.
\cite{busti2015cosmography} with the use of cosmographical analysis.
Capozziello \emph{et al}. provided cosmography as a non-predictive
phenomenon when the redshift parameter becomes $z\approx1$. They
used the pad$\acute{e}$ approximations for the fifth order and
resulted the divergence of data at the higher levels of the
approximations. Lobo \emph{et al}. \cite{lobo2022dynamical}
evaluated the dynamics of the redshift drift. They used the
expanding $FLRW$ universe to produce a general matter and low
redshift model with the use of different variables. However, the
 cross-correlation of  large-scale quasars can be used and translated
  with the CMB and \textbf{BAO} scale data to produce the best for Hubble
parameter $H(z)$ and angular diametric distance $S_{A}$. Also, the
cosmic chronometers approach can be done to predict the model
independent $H(z)$ measurements which have been extensively used for
cosmological applications
\cite{moresco20166,hu2021measuring,wang2022standardized}. The low
redshift data set with the inclusion of the megamasers and
chronometers had been presented by Krishnan and others
\cite{krishnan2020there}. They result that the Hubble constant
$\mathcal{H}_0$, showed descending behavior with the redshift and
having non-zero slop when fitted on the line by statistical means.
Font \emph{et al}. \cite{font2014quasar} studied correlation
technique for quasars by using Lya absorption and produced the best
line of fit for Planck's data. They generated different results on
the measurements of the Hubble parameter and the angular distance.
One important thing is to develop such a cosmic Hubble parameter
that comes from early to late span in such a way that it changes
from a low to a high value. The Gaussian method helped to predict
but provided a non-transitional behavior for both $\Lambda$ and
$\omega$ epochs. The null energy condition also proved to be an
important restriction for the cut-off model, when compared with
Hubble parameter data \cite{hu2022revealing}. King \emph{et al}.
\cite{king2014high} studied the future approximations of the
redshift by the inclusion of dark energy. They tested the equation
of state by the linear parametrization technique. Hu \emph{et al}.
\cite{hu2022revealing} reported different values of the Hubble
constant by the Gaussian method. Their research produced an
effective reduction in the Hubble crisis and proposed the
non-transitional behavior of the Hubble constant. Different dark
energy models respective to holography and agegraphy had been
conducted by Zhang \emph{et al}. \cite{zhang2013cosmological}. They
produced different energy conditions for different red shift values
and resulted in an effective role of energy conditions for different
cosmic ages. 

In this article, we implemented a functional form of the Hubble parameter
that evolves periodically with cosmic time $t$ and investigate the bouncing nature of the universe in
$f(\mathcal{G},\mathcal{T}_{\alpha \beta} \mathcal{T}^{\alpha
\beta})$ gravity using a flat $\mathcal{FLRW}$ peacetime. This
analysis of the bouncing universe involves one of the most important
forms of $EoS$ parameter proposed in the literature
\cite{babichev2004dark,haro2015gravitational,bacalhau2018consistent}.
The outline is given as: Sect.$\textbf{2}$ provides a brief
introduction to $f(\mathcal{G},\mathcal{T}_{\alpha \beta}
\mathcal{T}^{\alpha \beta})$ gravity with the necessary formalism of
$\mathcal{FLRW}$ metric and modified field equations. Sect.$\textbf{3}$
builds the Hubble parameter as a bouncing solution for the produced field
equations. The cosmographic parameters are
also evaluated in this section. We provide the mathematical expressions
of energy density and matter pressure for the assumed $EoS$ parameter
form in Sect.$\textbf{4}$. The energy conditions are also formulated in
the same fashion. Detailed graphical profiles of energy conditions are
represented in the same section to discuss the evolution of the universe
under the influence of
restricted free parameters. Finally, the concluding remarks are made in
Sect.$\textbf{5}$.

\section{$f(\mathcal{G},\mathcal{T}_{\alpha \beta} \mathcal{T}^{\alpha
\beta})$ Formalism}

The modified action for the $f(\mathcal{G},\mathcal{T}_{\alpha
\beta} \mathcal{T}^{\alpha \beta})$ gravity theory is defined as
\cite{yousaf2022f}
\begin{equation}\label{1}
\mathbb{A}_{f(\mathcal{G},\mathcal{T}_{\alpha \beta}
\mathcal{T}^{\alpha \beta})}=
\frac{\sqrt{-g}}{2\kappa^{2}}\left(\int
d^{4}x[f(\mathcal{G},\mathcal{T}_{\alpha \beta} \mathcal{T}^{\alpha
\beta})+\mathcal{R}] +\int d^{4}x \mathcal{L}_{m}\right),
\end{equation}
where $\mathcal{R}$ and $\mathcal{G}$ symbolize the Ricci scalar and
the Gauss-Bonnet scalar terms, respectively and are provided as
\begin{equation}\label{2}
\mathcal{R}\equiv g_{\alpha
\beta}\mathcal{R}^{\alpha\beta},~~~~\mathcal{G} \equiv
\mathcal{R}_{\xi\zeta\alpha\beta}\mathcal{R}^{\xi\zeta\alpha\beta}-4\mathcal{R}
_{\xi\zeta}\mathcal{R}^{\xi\zeta}+\mathcal{R}^2,
\end{equation}
and $\kappa^2$ = $8 \pi$G (G be the gravitational constant) and $\mathcal{L}_{m}=-p$. Also, the term $g$ implies the trace of the metric tensor $g_{\alpha \beta}$ with $T_{\alpha\beta}$, $R_{\xi\zeta\alpha\beta}$ and $R_{\alpha\beta}$ indicate the stress energy-momentum tensor, the Riemannian tensor, and the Ricci tensor respectively. The expression for $\mathcal{T}_{\alpha\beta}$ is given as
\begin{equation}\label{3}
\mathcal{T}_{\alpha\beta}=\frac{-2}{\sqrt{-g}}\frac{\delta(\sqrt{-g}\mathcal{L}_m)}{\delta
g^{\alpha\beta}}.
\end{equation}
Equation \eqref{3} yields the following expression, due the
dependency of the matter Lagrangian $\mathcal{L}_m$ on
$g_{\alpha\beta}$ components
\begin{equation}\label{4}
\mathcal{T}_{\alpha\beta}=g_{\alpha\beta}\mathcal{L}_m-\frac{2\partial\mathcal{L}_m}{\partial
g^{\alpha\beta}}.
\end{equation}
Now, by taking the variation of Eq.\eqref{1} with respect to the
term $g_{\alpha\beta}$, we get the following field equations for the
$f(\mathcal{G},\mathcal{T}_{\alpha \beta} \mathcal{T}^{\alpha
\beta})$ theory as
\begin{equation}\label{5}
\mathcal{R}_{\alpha\beta}-\frac{1}{2}\mathcal{R}
g_{\alpha\beta}=\mathcal{T}_{\alpha\beta}^{\emph{eff}},
\end{equation}
where the term $\mathcal{T}_{\alpha\beta}^{\emph{eff}}$ takes the following form
\begin{eqnarray}\nonumber
\mathcal{T}_{\alpha\beta}^{\emph{eff}}&=&\kappa^{2}\mathcal{T}_{\alpha\beta}-\Theta_{\alpha\beta}f_{\mathcal{T}^2}(\mathcal{G},\mathcal{T}^{2})
+\frac{1}{2}g_{\alpha\beta}f(\mathcal{G},\mathcal{T}^2)-(2\mathcal{R}\mathcal{R}_{\alpha\beta}-4\mathcal{R}_{\alpha}^{\varepsilon}\mathcal{R}_{\varepsilon\beta}
-4\mathcal{R}_{\alpha\varepsilon\beta\eta}\mathcal{R}^{\varepsilon\eta}\\\nonumber
&&+2\mathcal{R}_{\alpha}^{\varepsilon\eta\delta}
\mathcal{R}_{\beta\varepsilon\eta\delta})f_{\mathcal{G}}(\mathcal{G},\mathcal{T}^2)-(2\mathcal{R}\nabla^{2}g_{\alpha\beta}-2\mathcal{R}\nabla_{\alpha}\nabla_{\beta}
-4\mathcal{R}_{\alpha\beta}\nabla^{2}-4g_{\alpha\beta}\mathcal{R}^{\varepsilon\eta}
\nabla_{\varepsilon}
\nabla_{\eta}\\\label{6}&&+4\mathcal{R}^{\varepsilon}_{\alpha}\nabla_{\beta}\nabla_{\varepsilon}
+4\nabla_{\varepsilon}\nabla_{\alpha}\mathcal{R}^{\varepsilon}_{\beta}
+4\mathcal{R}_{\alpha\varepsilon\beta\eta}\nabla^{\varepsilon}\nabla^{\eta})
f_{\mathcal{G}}(\mathcal{G},\mathcal{T}^2),
\end{eqnarray}
where,
\begin{equation}\label{7}
\Theta_{\alpha \beta}\equiv\frac{\delta(\mathcal{T}_{\mu\nu}
\mathcal{T}^{\mu\nu})}{\delta g^{\alpha\beta}}
=2\mathcal{T}_{\alpha}^{\xi}\mathcal{T}_{\beta\xi}-\mathcal{T}
\mathcal{T}_{\alpha\beta}-4\mathcal{T}^{\mu\nu} \frac{\partial^2
\mathcal{L}_m}{\partial g^{\alpha\beta}g^{\mu\nu}}-2\mathcal{L}_m
(\mathcal{T}_{\alpha\beta}-\frac{1}{2}\mathcal{T}g_{\alpha\beta})
\end{equation}
\begin{equation}\label{8}
\mathcal{T}^2=\mathcal{T}_{\alpha
\beta} \mathcal{T}^{\alpha \beta},~~~~~\nabla^{2}=\nabla_{\alpha}\nabla^{\alpha}
\end{equation}
The  terms $f_{\mathcal{G}}(\mathcal{G},\mathcal{T}^2)$ and
$f_{\mathcal{T}^2}(\mathcal{G},\mathcal{T}^2)$ used above are
defined as
\begin{equation}\label{9}
~~f_{\mathcal{G}}(\mathcal{G},\mathcal{T}^2)\equiv\frac{df(\mathcal{G},\mathcal{T}^2)}{d\mathcal{G}},~~
~~and~~f_{\mathcal{T}^2}(\mathcal{G},\mathcal{T}^2)\equiv\frac{df(\mathcal{G},\mathcal{T}^2)}{d\mathcal{T}^2}.
\end{equation}
The trace of the above-defined field equations is produced as
\begin{equation}\label{10}
\mathcal{T}-\Theta f_{\mathcal{T}^2}(\mathcal{G},\mathcal{T}^2)+2
\mathcal{G}f_{\mathcal{G}}(\mathcal{G},\mathcal{T}^2)-2\mathcal{R}
\nabla^2f_{\mathcal{G}}(\mathcal{G},\mathcal{T}^2)+4
\mathcal{R}_{\alpha\beta}\nabla^{\alpha}\nabla^{\beta}
f_{\mathcal{G}}(\mathcal{G},\mathcal{T}^2)=0.
\end{equation}
Equation \eqref{10} shows the non-conversed situation of the stress energy-momentum tensor.
 Also, the properties of $\mathcal{GR}$ can be recovered for
 $f(\mathcal{G},\mathcal{T}^2)=0$. Similarly if we put
 $f(\mathcal{G},\mathcal{T}^2)=f(\mathcal{G})$,
we get the properties of $f(\mathcal{G})$ gravity.

Now, as we are concerned to study the bouncing nature of the
universe, so we consider the fluid distribution to be perfect
throughout the cosmic
 evolution. For this, we take
\begin{equation}
\mathcal{T}_{\alpha\beta}=(\rho+p)V_\alpha
V_\beta-pg_{\alpha\beta},\label{11}
\end{equation}
here, the four-vector velocity is
defined by $V^\beta$ with
\begin{equation}\label{12}
V^\beta=(1,0,0,0),~~ V^\beta V_\beta=1~,~
V^\beta\nabla_\zeta V_\zeta=0.
\end{equation}
In addition, $\rho$ defines the energy density part and $p$
defines the pressure part of the stress energy-momentum tensor. Also the geometric
 background considered to be in a $\mathcal{FLRW}$ space time\cite{melia2022friedmann}, so it implies
\begin{equation}
ds^2=dt^2-a^2(t)\Sigma_{i}dx_{i}^2,~~~~~~~~~~~i=1,2,3.\label{13}
\end{equation}
The metric component $a(t)$ symbolizes the scale factor, that
contributes to the Hubble parameter as
$\mathcal{H}=\frac{\dot{a}(t)}{a(t)}$. Using Eq.\eqref{13} and
Eq.\eqref{7} in Eq.\eqref{5}, we get the following field equations
\begin{equation}\label{14}
6\left(\frac{\dot{a}}{a}\right)^{2}-24\left(\frac{\dot{a}}{a}\right)^{3}\dot{f_\mathcal{G}}
+24\left(\frac{\ddot{a}}{a}\right)\left(\frac{\dot{a}}{a}\right)^{2}f_\mathcal{G}-f-2(\rho^2
+3p^2+4\rho p)f_{\mathcal{T}^2}=2\rho\kappa^2,
\end{equation}
\begin{equation}\label{15}
-2\left(2\frac{\ddot{a}}{a}+
\left(\frac{\dot{a}}{a}\right)^2\right)+16\left(\frac{\ddot{a}\dot{a}}{a^2}\right)\dot{f_G}+8\left(\frac{\dot{a}}{a}\right)^{2}
\ddot{f_G}-24\left(\frac{\ddot{a}}{a}\right)\left(\frac{\dot{a}}{a}\right)^{2}f_{G}
+f=2p\kappa^{2}.
\end{equation}
To draw the conclusions on the field equations, we just need some
functional form of  $f(\mathcal{G},\mathcal{T}^2)$. As, there are
many functional forms regarding the interaction of matter with the
curvature terms, in order to deal with the issues of cosmic
evolution. Various coupling models can be used to evaluate the
formations of both energy density and matter pressure, like one can
take $f(\mathcal{G}, \mathcal{T}^2) = \mathcal{G} +
2f(\mathcal{T}^2)$ that may help to provide an analysis about
$\Lambda CDM$ epoch. However, the other choice is $f(\mathcal{G},
\mathcal{T}^2) = f_1(\mathcal{G}) + f_2(\mathcal{T}^2)$ that may be
worked as a correction to $f(\mathcal{G})$ gravity theory because of
$f_2(\mathcal{T}^2)$. Similar forms have been explored in
\cite{yousaf2016causes,shamir2021bouncing} and provided some
distinct results due to the direct minimal curvature matter
coupling. Also, $f(\mathcal{G},\mathcal{T}^2) = f_1(\mathcal{G}) +
f_2(\mathcal{G})f_3(\mathcal{T}^2)$ can be taken because of an
explicit non-minimally coupling nature between geometric parameters
and matter variables \cite{nojiri2017modified}. So, we considered
the following form to produce the validating results.
\begin{equation}\label{16}
f(\mathcal{G},\mathcal{T}^2)=f_1(\mathcal{G})+f_2(\mathcal{T}^2).
\end{equation}
To produce a bouncing universe, we need some functional forms of
$f_1$ and $f_2$ that not only describe the accelerating expansion of
the universe but also explain inflation to a great extent. For this,
the higher power curvature terms perform well to eliminate such
issues. Elizalde \cite{elizalde2020cosmological} introduced the
power forms of the curvature scalar as $\eta \mathcal{R}^n$
($n\geq1$) and produced the cosmological dynamics, so we consider
the specific form of the $f_1$ as the quadratic power model, so
\begin{equation}\label{17}
f_1(\mathcal{G})= \mathcal{G}+\alpha \mathcal{G}^2.
\end{equation}
Also, we take $\chi_2$ as
\begin{equation}\label{18}
f_2(\mathcal{T}^2)=2\lambda \mathcal{T}^2.
\end{equation}
So, by using Eq.s \eqref{17} and \eqref{18} in the field equations,
we get
\begin{eqnarray}\label{19}
6 \mathcal{H}^2 -48\alpha \mathcal{H}^3 \mathcal{G}
\dot{\mathcal{G}}+\alpha \mathcal{G}^2=2 \kappa^2
\rho+6\lambda\rho^2+18\lambda p^2 + 16 \lambda\rho p,
\end{eqnarray}
and
\begin{eqnarray}\label{20}
-2(2\dot{\mathcal{H}}+3
\mathcal{H}^2)+32(\dot{\mathcal{\mathcal{H}}}+\mathcal{H}^2)\alpha
\mathcal{G} \dot{\mathcal{G}}+16\alpha
\mathcal{H}^2(\dot{\mathcal{G}}^2+\mathcal{G}\ddot{\mathcal{G}})-\alpha
\mathcal{G}^2= 2\kappa^2 p - 2 \lambda\rho^2 -6\lambda p^2.
\end{eqnarray}
In order to reduce the complexity of the Eq.\eqref{19} and
Eq.\eqref{20}, we utilize $p=\omega \rho$, as the \emph{EoS} used in
\cite{babichev2004dark,haro2015gravitational,bacalhau2018consistent}.
So we get the relations,
\begin{eqnarray}\label{21}
(3\lambda+9\lambda\omega^2 +8 \lambda\omega)\rho^2 +\kappa^2 \rho-(3
\mathcal{H}^2 -24\alpha \mathcal{H}^3 G
\dot{\mathcal{G}}+\frac{\alpha}{2} \mathcal{G}^2)=0
\end{eqnarray}
and
\begin{eqnarray}\label{22}
(-\frac{\lambda}{\omega^2}-3\lambda)p^2 +\kappa^2 p
+((2\dot{\mathcal{H}}+3H^2)-16(\dot{\mathcal{H}}+\mathcal{H}^2)\alpha
\mathcal{G} \dot{\mathcal{G}}-8\alpha
\mathcal{H}^2(\dot{\mathcal{G}}^2+\mathcal{G}\ddot{\mathcal{G}})+\frac{\alpha}{2}
\mathcal{G}^2)=0.
\end{eqnarray}
where,
$\mathcal{G}=24\mathcal{H}^2(\dot{\mathcal{H}}+\mathcal{H}^2)$. Yousaf and his collaborators checked the stability of
cosmic models in various modified gravity theories \cite{sharif2015instability,bhatti2018existence,yousaf2020definition,nasir2022influence}.

\section{Hubble Parameter and Cosmography}

\begin{figure}
\epsfig{file=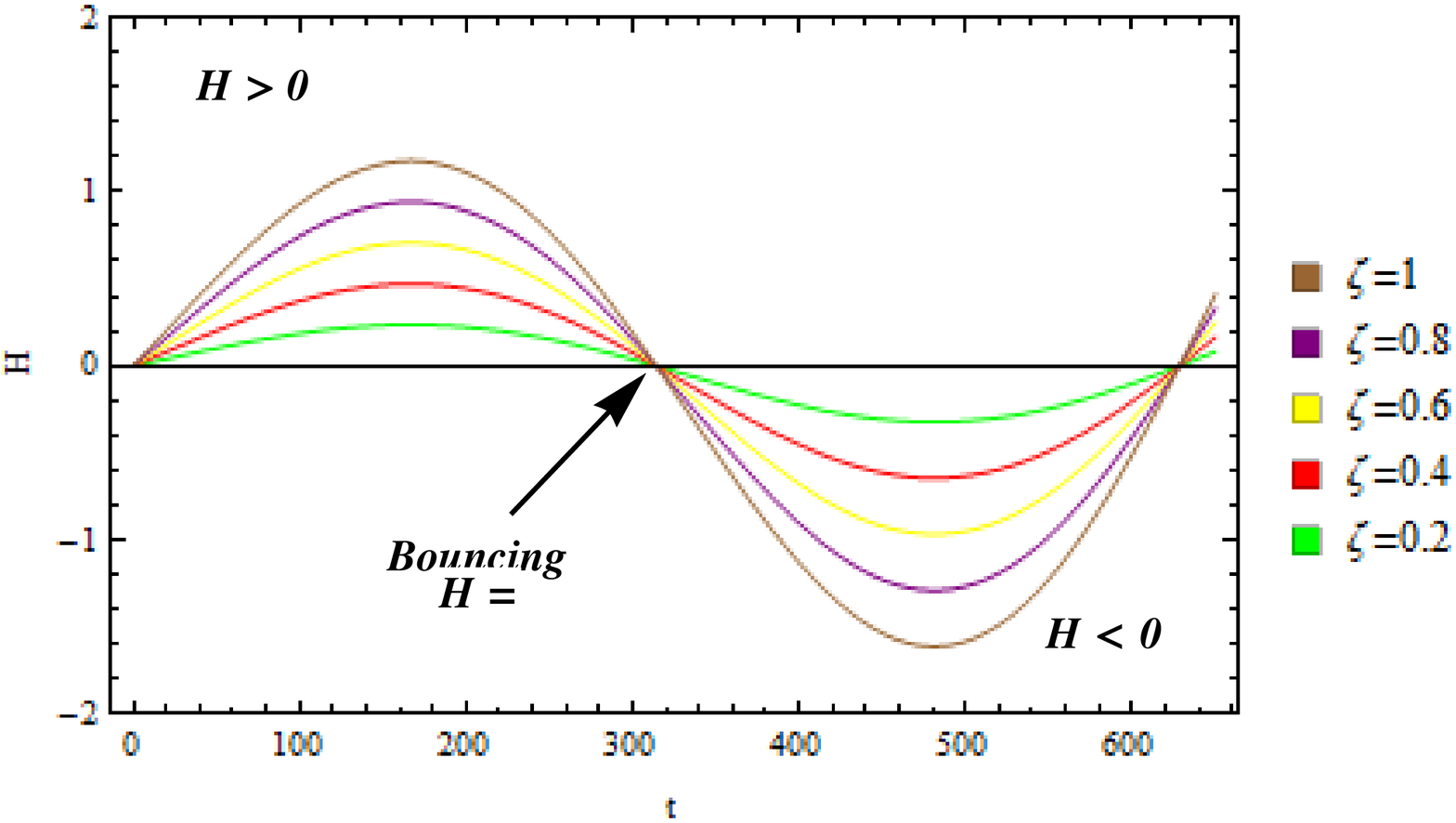,width=.52\linewidth}
\epsfig{file=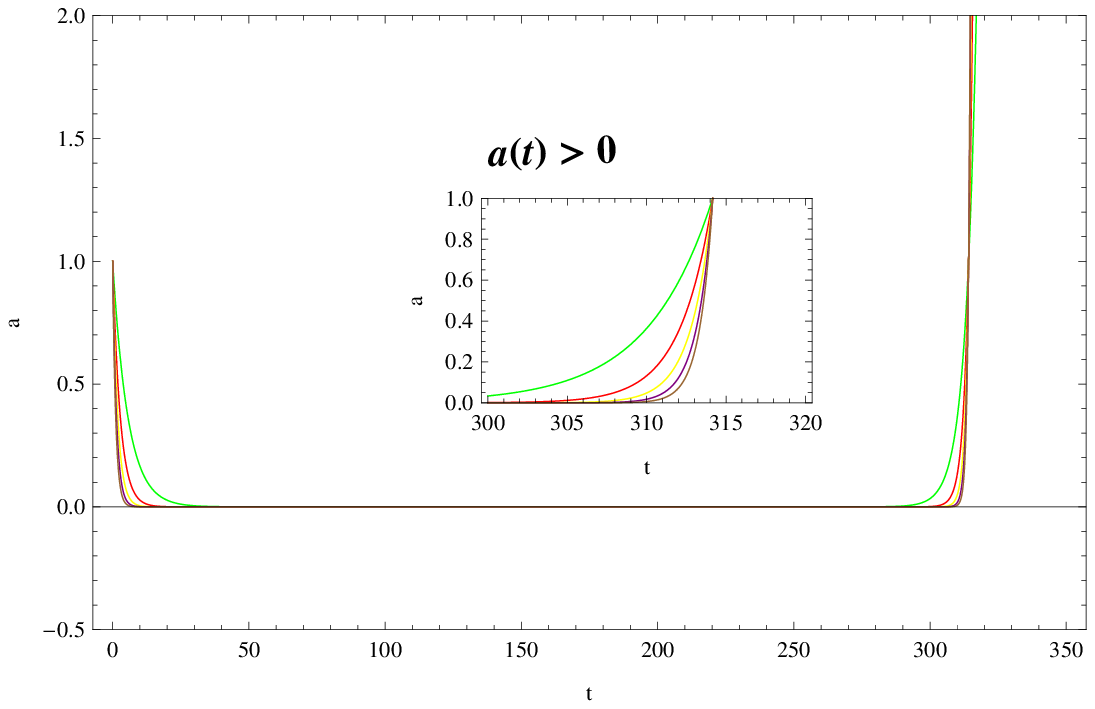,width=.46\linewidth} \caption{The illustrations
of Hubble parameter and scale factor with fixed values of $\varphi
=0.001$ and $\phi =0.01$.}\label{F1}
\end{figure}

\begin{figure}
\epsfig{file=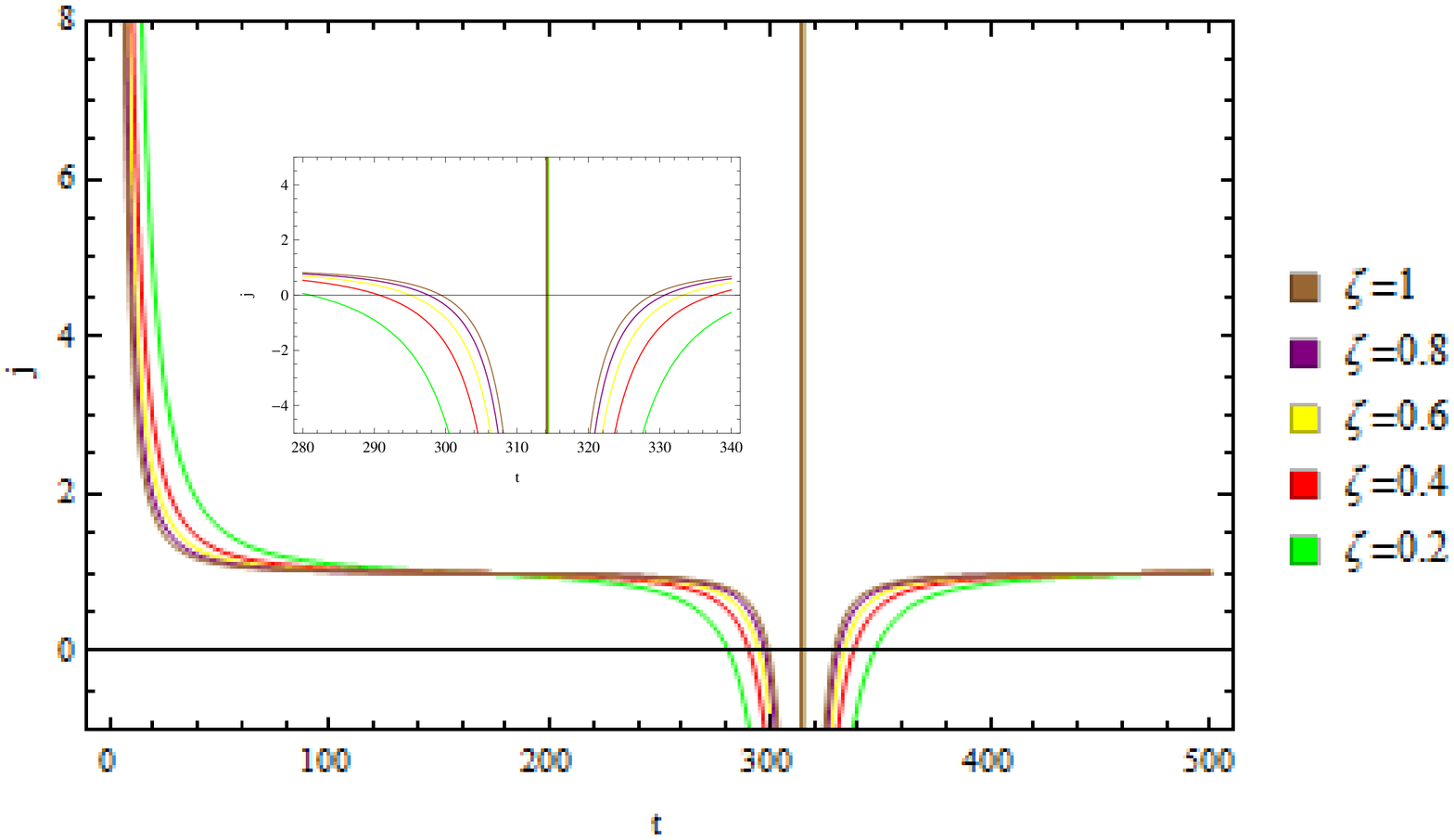,width=.52\linewidth}
\epsfig{file=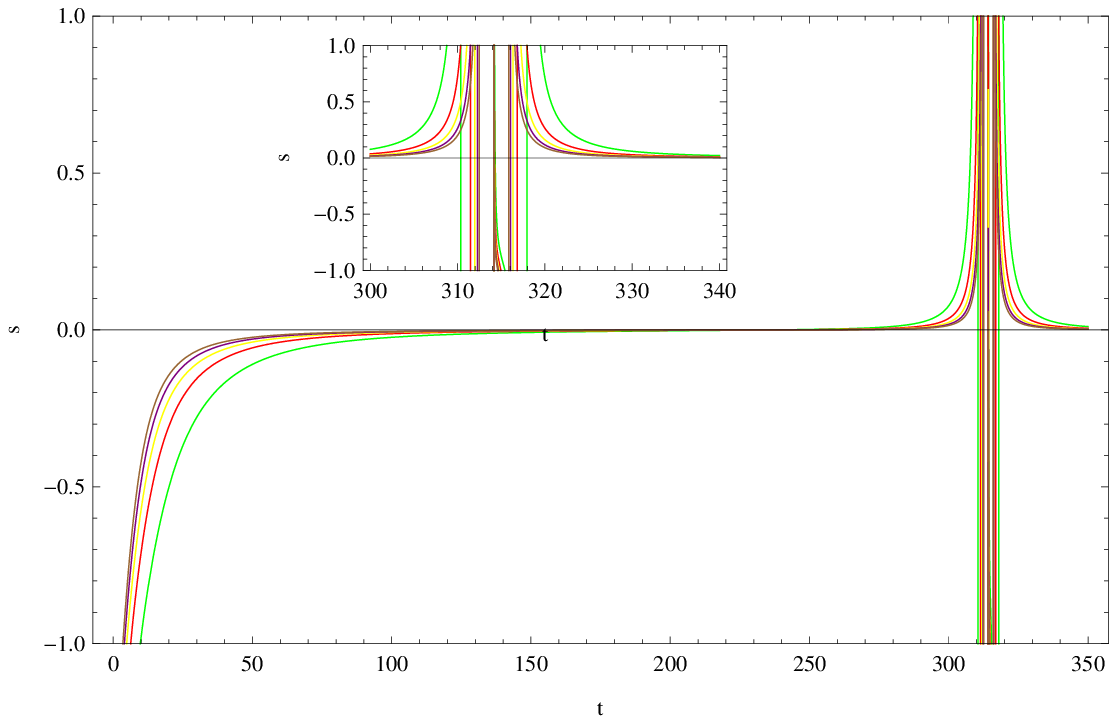,width=.46\linewidth} \caption{The illustration of
jerk and snap factors with fixed values of $\varphi =0.001$ and
$\phi =0.01$.}\label{F2}
\end{figure}

This section mainly focuses on describing the evolutionary behavior of these above-described dynamical terms. Hence, we consider a trigonometric form of the $\mathcal{H}(t)$ which feasibly provides a bounce solution \cite{shamir2021abouncing,elizalde2020cosmological},
as follows
\begin{equation}\label{23}
\mathcal{H}(t)=\zeta  \sin ( \phi t) h(t).
\end{equation}
This parameterized form of $\mathcal{H}(t)$ includes $\zeta$ and
$\phi$, which are considered to be constants here. The choice of
$h(t)$ depends on the periodic values of the function $\sin ( \phi
t)$, so the form of $h(t)$ can be chosen as periodic, that
cooperates with the non-vanishing values of the above trigonometric
function. This artificial approach of choosing such an ansatz can be
considered as a numerical analysis of making the bouncing solution.
One interesting feature is possessed by the term $\zeta$, which can
work well as a phase changer for the value of $\mathcal{H}(t)$. We
consider $h(t)$ as
\begin{equation}\label{24}
h(t)=\exp (\varphi t),
\end{equation}
where $\varphi$ acts as a constant. Finally, we have
\begin{equation}\label{25}
\mathcal{H}(t)=\zeta  \sin ( \phi t) \exp(\varphi t).
\end{equation}
This functional form of the Hubble parameter is helpful to study
cosmic evolutionary expansion and contraction. This form of the Hubble parameter gives us the bounce at $t=313$, depending upon the values of $\varphi =0.001$ and $\phi =0.01$ provided in Fig.\ref{F1}. We have restricted the values of $\mathcal{H}(t)$ in the positive era of time. The basic scale factor form for this parameterized Hubble parameter becomes
\begin{equation}\label{26}
a(t)=\exp \left(\frac{\zeta  \exp ( \varphi  t) (\varphi  \sin (
\phi t)-\phi  \sin ( \phi t))}{\varphi ^2+\phi ^2}\right).
\end{equation}
Similarly, the set of dynamical parameters that are derived from the
Taylor series expansion of the scale factor is termed as
cosmographic factors. These factors helped to obtain the
cosmological concordance with the assumptions of the universal
homogeneity and isotropy on large cosmic scales \cite{busti2015cosmography,hu2022high}. These include deceleration, jerk and snap parameters. These factors allow us to check the
compatibility of the scale factor and the Hubble parameter. The negative value of the deceleration parameter $q$ describes the accelerated expansion of the universe. Similarly, jerk $j$ and snap $s$ determine the expansion rate of the toy universe model. The
mathematical interpretation for these cosmography elements are
defined as
\begin{equation}\label{27}
q=-\frac{1}{\mathcal{H}^2}\frac{1}{a}\frac{d^2
a}{dt^2}=-1-\frac{1}{\zeta }(e^{- \varphi t} \csc ( \phi t) (\phi
\cot ( \phi t )+\varphi )),
\end{equation}
\begin{eqnarray}\nonumber
j=\frac{1}{\mathcal{H}^3}\frac{1}{a}\frac{d^3
a}{dt^3}&=&1+\frac{1}{\zeta ^2}(e^{-2  \varphi t} \csc ( \phi t) (3
\zeta  e^{ \varphi t} (\phi  \cot ( \phi t)+\varphi
)\\\label{28}&&+\csc ( \phi t) (2 \varphi  \phi  \cot ( \phi
t)+\varphi ^2-\phi ^2))),
\end{eqnarray}
and
\begin{eqnarray}\nonumber
s=\frac{1}{\mathcal{H}^4}\frac{1}{a}\frac{d^4 a}{dt^4}=-\frac{1}{3
\zeta  (3 \zeta  e^{ \varphi t}+2 \csc ( \phi t) (\phi  \cot ( \phi
t)+\varphi ))}(2 e^{- \varphi t} \csc ( \phi t) (\csc ( \phi
t)\\\label{29}(3 \zeta \varphi  e^{ \varphi t} \sin ( \phi
t)+\varphi ^2-\phi ^2)+\phi  \cot ( \phi t) (3 \zeta e^{ \varphi
t}+2 \varphi \csc ( \phi t)))).
\end{eqnarray}
Fig.\ref{F1} shows the progression of the Hubble (left
panel) and scale parameters (right panel) along the positive time
axis. Similarly, the development of jerk (left panel) and
snap factors (right panel) are provided in the fig.\ref{F2}. The
evolution of the deceleration parameter towards the negative value
i.e, $q\rightarrow-1$, before the bouncing point, provided in
fig.\ref{f6}, shows the accelerating universe.

\section{Energy Conditions under the EoS Parameter}
\begin{figure}
\epsfig{file=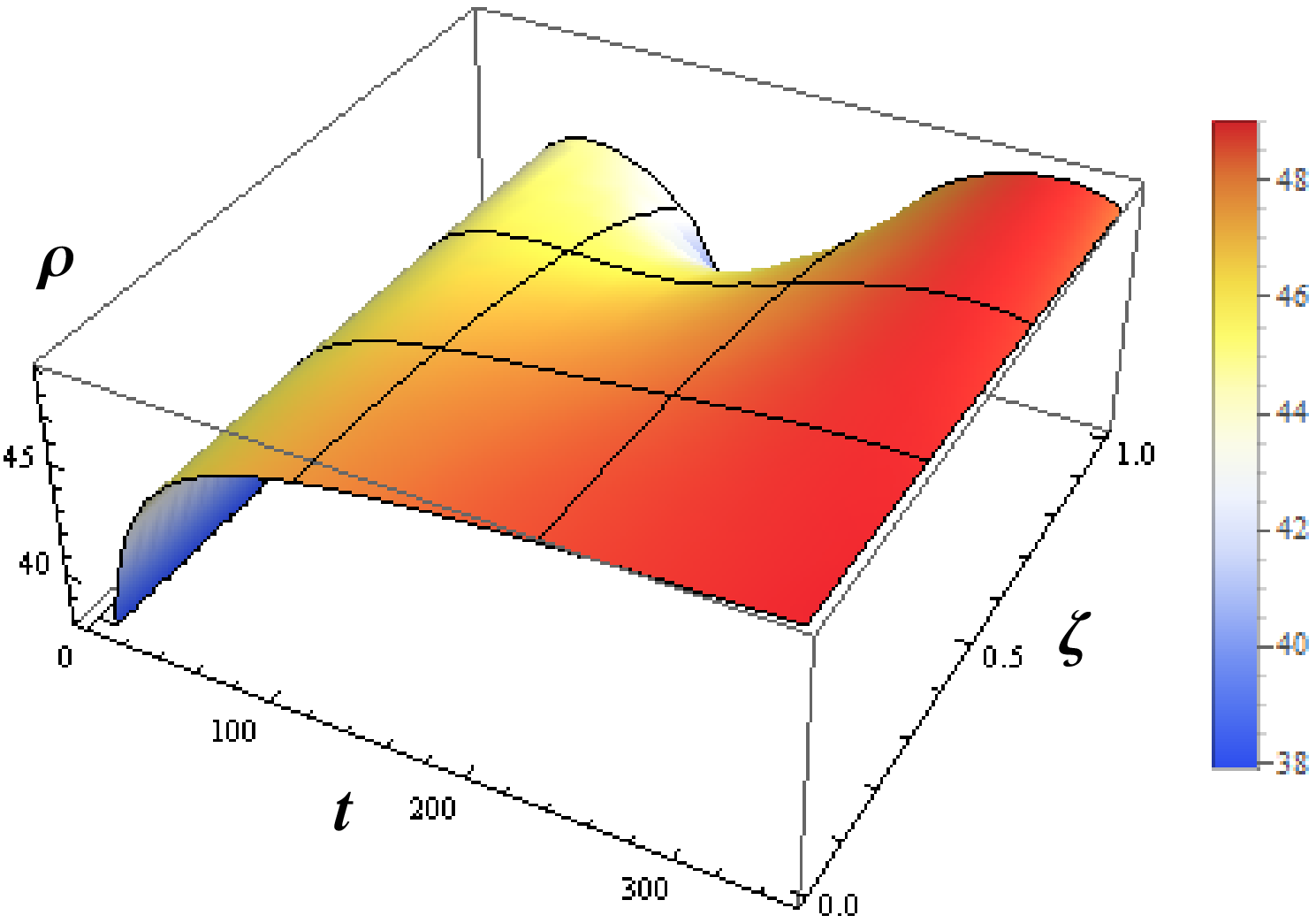,width=.52\linewidth}
\epsfig{file=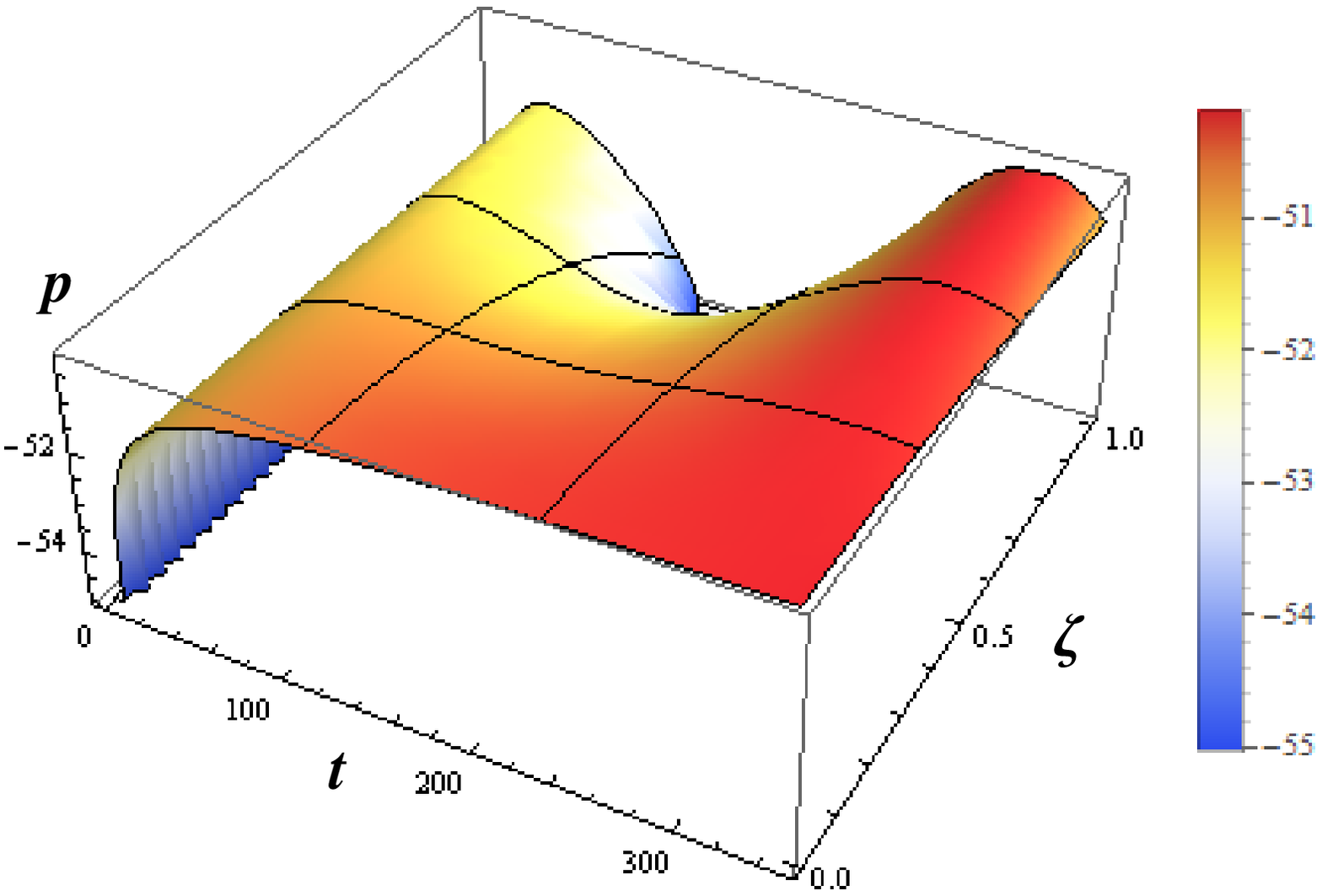,width=.52\linewidth} \caption{The illustration of
energy density  and matter pressure with fixed values of $\alpha
=0.005$, $k=0.5$, $\varphi =0.001$, $\epsilon =0.001$, $\phi =0.01$,
$\kappa =1$ and $\lambda =-0.005$.}\label{f3}
\end{figure}

\begin{figure}
\epsfig{file=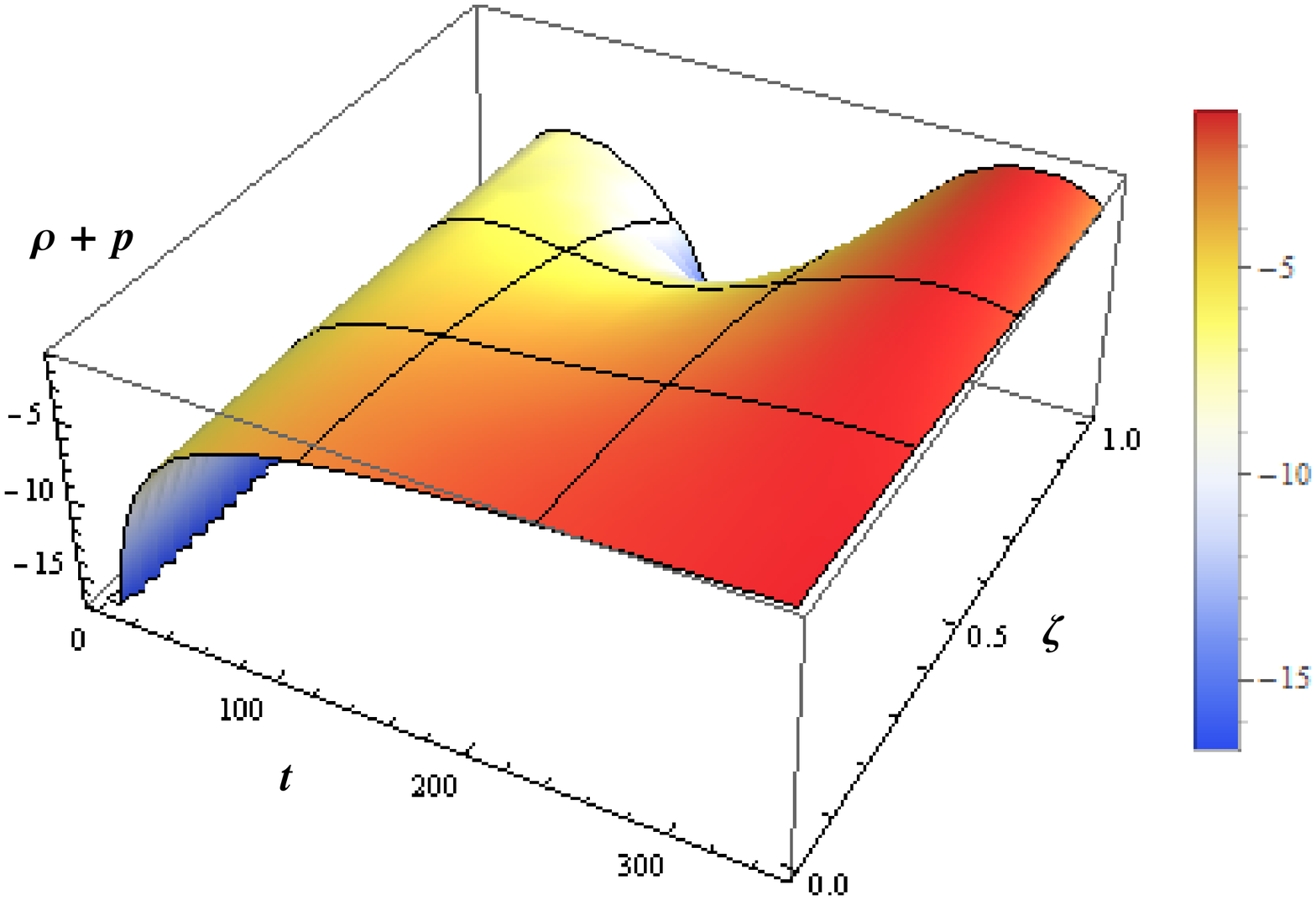,width=.52\linewidth}
\epsfig{file=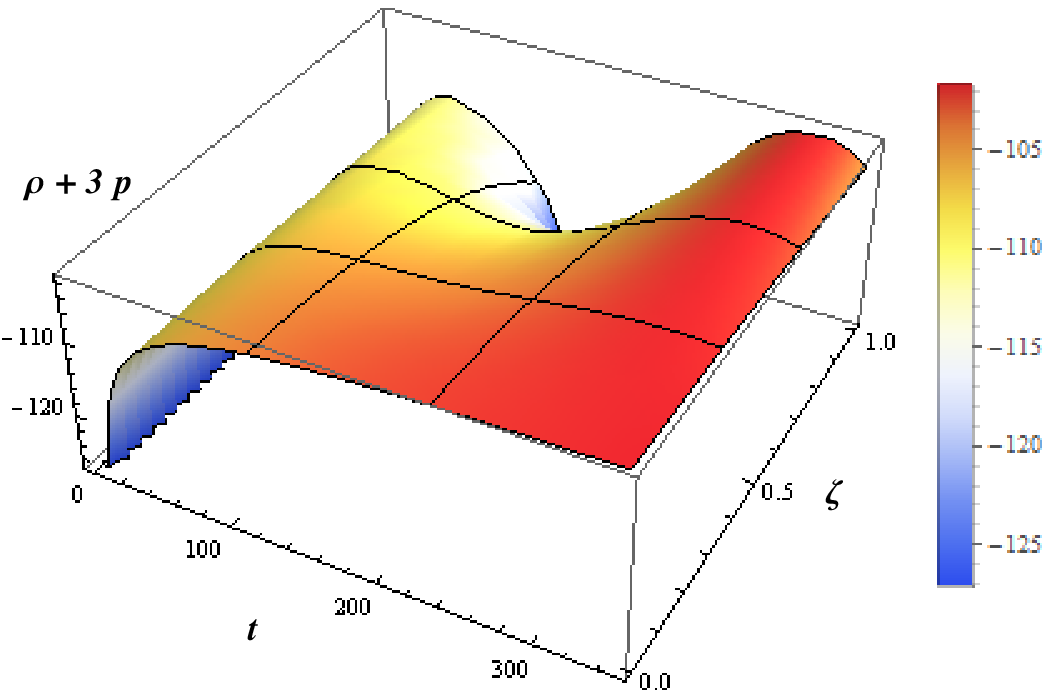,width=.52\linewidth} \caption{The illustration
of $\mathcal{NEC}$ and $\mathcal{SEC}$ with fixed values of $\alpha
=0.005$, $k=0.5$, $\varphi =0.001$, $\epsilon =0.001$, $\phi =0.01$,
$\kappa =1$ and $\lambda =-0.005$.}\label{f4}
\end{figure}

\begin{figure}
\epsfig{file=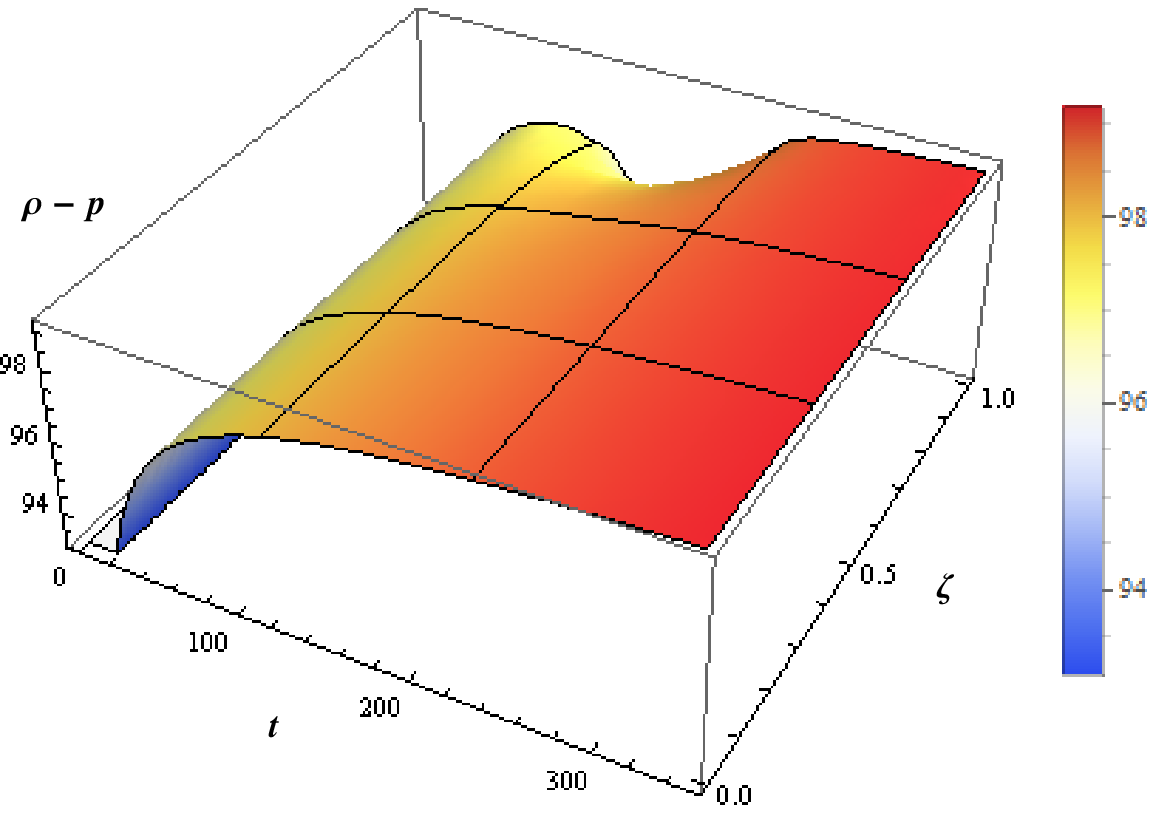,width=.52\linewidth}
\epsfig{file=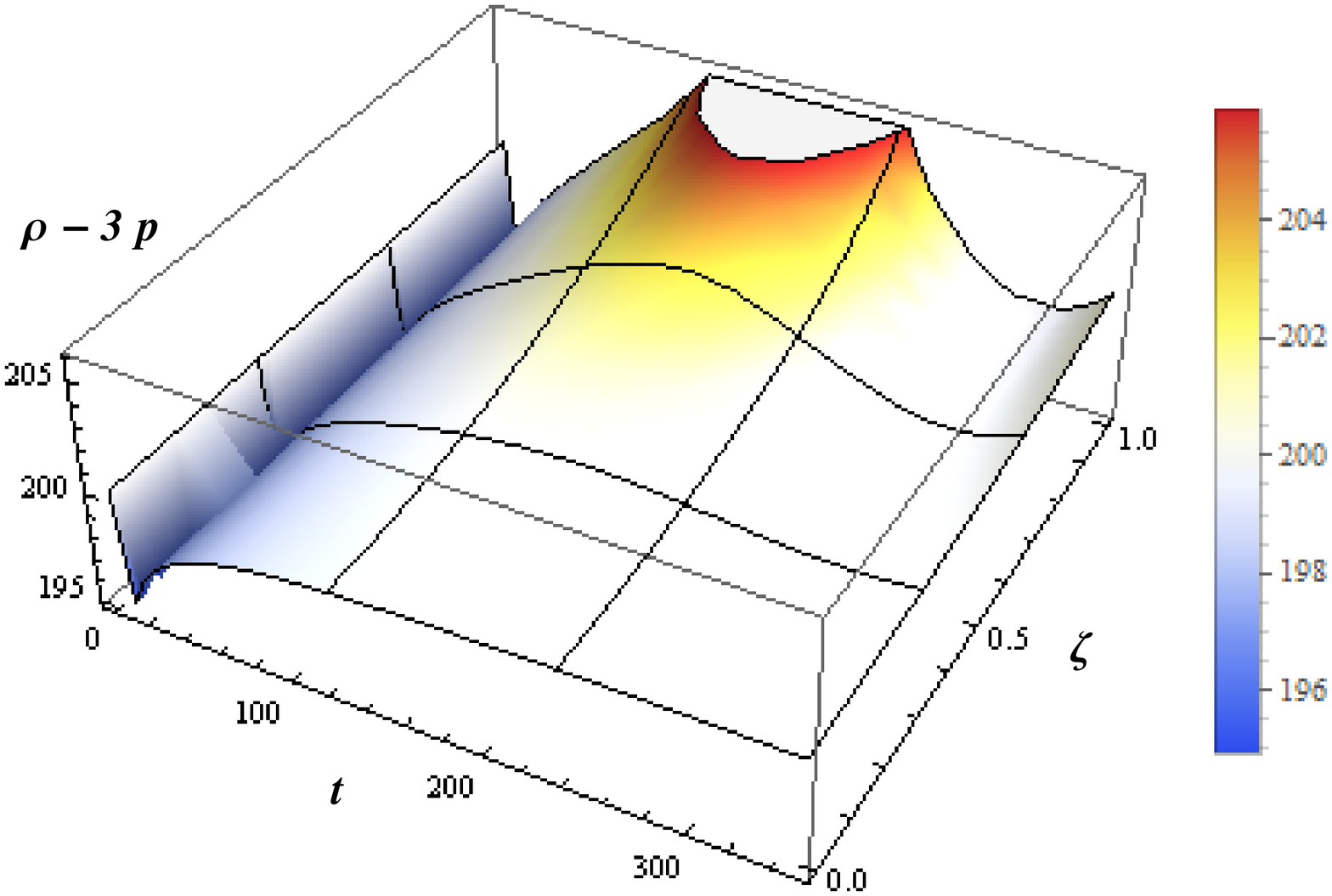,width=.52\linewidth} \caption{The illustration
of $\mathcal{DEC}$ and $\mathcal{TEC}$ with fixed values of $\alpha
=0.005$, $k=0.5$, $\varphi =0.001$, $\epsilon =0.001$, $\phi =0.01$,
$\kappa =1$ and $\lambda =-0.005$.}\label{f5}
\end{figure}

\begin{figure}
\epsfig{file=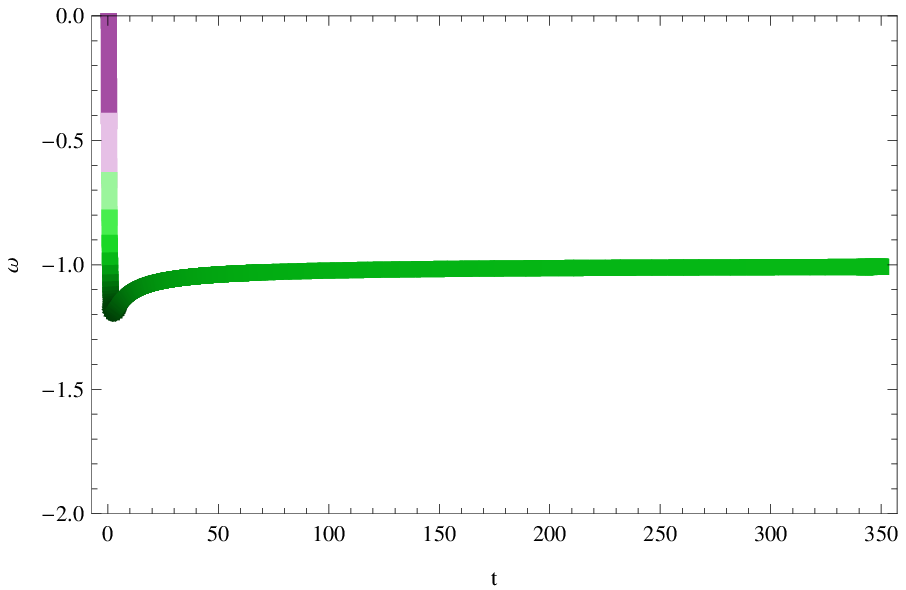,width=.52\linewidth}\epsfig{file=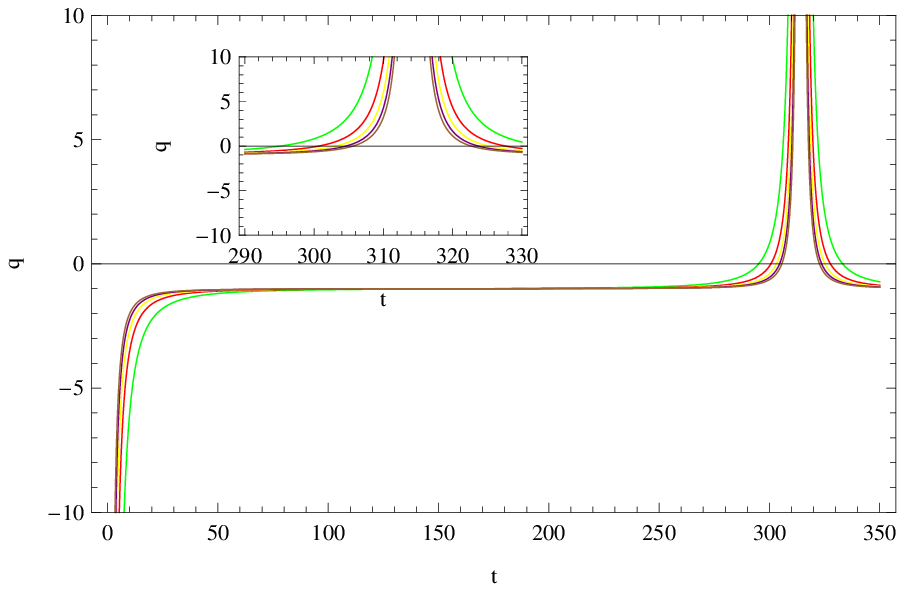,width=.52\linewidth}
\caption{The illustration of $\emph{EoS}$ and deceleration
parameters with fixed values of $k=0.5$ and $\epsilon
=0.001$.}\label{f6}
\end{figure}

For a specific cosmology model, energy conditions play an important
role to make its validation for the restricted free parameters.
These energy conditions help to maintain the specifications of the
certain cosmic model
\cite{hawking1973large,visser1997energy,nojiri2003effective,bertolami2009energy
,balart2014regular}.
Similarly, these energy conditions also work for the bouncing
cosmology and provide a reasonable approach to validate the
procedure for our toy bouncing model. These conditions are described
as
\begin{itemize}
  \item Dominant energy condition ($\mathcal{DEC}$)$\Leftrightarrow$ $\rho\geq0$ , $\rho\pm
  p\geq0$.
  \item Strong energy condition ($\mathcal{SEC}$)$\Leftrightarrow$ $\rho+3p\geq0$ ,
  $\rho+p\geq0$.
  \item Weak energy condition ($\mathcal{WEC}$)$\Leftrightarrow$ $\rho\geq0$ ,
  $\rho+p\geq0$.
  \item Null energy condition ($\mathcal{NEC}$)$\Leftrightarrow$
  $\rho+p\geq0$.
  \item Trace energy condition ($\mathcal{TEC}$)$\Leftrightarrow$ $\rho-3p\geq0.$

\end{itemize}
The positivity of $\mathcal{DEC}$, $\mathcal{SEC}$ and
$\mathcal{WEC}$ passes on the validity and necessity of the bouncing
concept. However, the violation of $\mathcal{NEC}$ has a major role.
This violation is different in the $\mathcal{GR}$ context.
Universal bouncing scenario is one of those ideas that
provides a chance to discuss the singularity-free universal
beginning. Many proposals in the literature suggested avoiding this
singularity through quantum aspects, but these don't have such
reliability to fit in the gravitational theory.  So, at this
point gravitational theories allow a specific mechanism to check
the validity of the bounce model and as well its own. Null energy
condition is one such tool to help achieve the task.
Also, it has been proved that in the context of $\mathcal{GR}$, the
violation of $\mathcal{NEC}$ is extremely difficult to be achieved
for local-field models. So, effective field theories provide a
chance to recognize the violation of the $\mathcal{NEC}$ and to
allow a non-singular bounce
\cite{larson2011seven,caldwell2002phantom,alam2004there,onemli2004quantum}.
One such effective field is $f(\mathcal{G},\mathcal{T}_{\alpha
\beta} \mathcal{T}^{\alpha \beta})$  theory that provides a chance
to study the quadratic nature of the energy terms i.e, energy
density and matter pressure\cite{yousaf2022non,yousaf2022f}. However,
it also allows getting a non-singular bounce for the assumed gravity
model form. For an excellent bouncing model, the value of
$\mathcal{H}(t)$ turns out to be
$\mathcal{\dot{H}}=-4G\rho\pi(1+\omega)>0$ for the formulation of
$\mathcal{GR}$. However, if the $\mathcal{NEC}$ gets violated, we
have the surety to get a bouncing scenario. To provide the
mathematical formulation of the energy conditions, we consider Eqs.
\eqref{21} and \eqref{22}. Also, the $EoS$ parameter in the negative
regime provides the present cosmic evolution
\cite{hogan2007unseen,corasaniti2004foundations,weller2003large} and
becomes favorable in the bouncing context with $\omega(t)\approx-1$.
However, bouncing cosmology provides the possible geodesic
evolution of the universe by avoiding the singularity along with the
resolution of the horizon problem, flatness problem, entropy problem
and many more \cite{ijjas2018bouncing}. For the modified gravity,
$EoS$ parameter enables us to study the universal dynamics. In this
study, we used $EoS$ parameter \cite{elizalde2020cosmological} to
obtain the possible chance of obtaining a bounce solution in
$f(\mathcal{G},\mathcal{T}^2)$  as
\begin{eqnarray}\label{30}
\omega(t)=-\frac{k \log (t+\epsilon )}{t}-1,
\end{eqnarray}
here $k$ is assumed to be a constant. This particular form of the
$EoS$ parameters allows us to study the contracting and expanding
behavior without involving the Hubble parameter as well as the scale
factor. Elizalde \emph{et al}.
\cite{elizalde2020cosmological} produces cosmological dynamics by
considering $\mathcal{R}^{2}$ gravity and logarithmic trace terms.
They checked the effects of the $\lambda$ parameter in the gravity model
$f(\mathcal{R}, \mathcal{T}) = \mathcal{R} + \lambda\mathcal{R}^2 +
2\beta \ln(\mathcal{T})$ along with the bouncing solution depending
on the two $EOS$ parameters. Our work first described the choice of
Hubble parameter and its effects on the dynamical field equations
and then involves the $EOS$ parameter. We only took one of the
$\omega(t)$ value, because this state factor after the bouncing
point remains negative and becomes $\omega(t)\approx-1$. Also, the
current cosmic expansion and $\Lambda-CDM$ can be verified by this
state factor. However, the dynamic properties are greatly
affected under the influence of this $EoS$ parameter form. Hence,
the general forms of the Eqs.\eqref{21} and \eqref{22}, under the
influence of Eq.\ref{30}, are presented as
\begin{eqnarray}\nonumber
\rho&=&-\frac{1}{2 \lambda (9 \omega ^2+8 \omega +3)}(\kappa
^2+(\kappa ^4-12 \zeta ^2 \lambda  (9 \omega ^2+8 \omega +3) e^{2
\varphi t } \sin ^2( \phi t) (2304 \alpha  \zeta ^7 e^{7  \varphi t}
\sin ^4( \phi t) \\\nonumber &&(\sin ( \phi t) (\zeta  e^{ \varphi
t} \sin ( \phi t)+\varphi )+\phi \cos ( \phi t)) (4 \zeta \varphi
e^{ \varphi t} \sin ^3( \phi t)+2 \zeta \phi e^{ \varphi t} \sin (2
 \phi t) \sin ( \phi t)\\\nonumber &-&(\phi ^2-3 \varphi ^2) \sin
^2( \phi t)+ 2 \phi ^2 \cos ^2( \phi t)+3 \varphi \phi \sin (2 \phi
t ))-96 \alpha \zeta ^4 e^{4  \varphi t} \sin ^2( \phi
t)\\\label{31} && (\sin ( \phi t) (\zeta e^{ \varphi t} \sin ( \phi
t)+\varphi )+\phi \cos ( \phi t))^2-1))^\frac{1}{2}
\end{eqnarray}
\begin{eqnarray}\nonumber
p&=&\frac{1}{2 (3 \lambda  \omega ^2+\lambda )}(\kappa ^2 \omega
^2+(\kappa ^4 \omega ^4+4 \zeta \omega ^2 (3 \lambda  \omega
^2+\lambda ) e^{ \varphi t} (18432 \alpha  \zeta ^9 e^{9 t \varphi }
(2 \varphi  (2 \varphi +1)-\phi ^2)\\\nonumber && \sin ^{10}(t\phi
t)+4608 \alpha \zeta ^8 e^{8 t \varphi } (\varphi ^2 (25 \varphi
+22)-(13 \varphi +2) \phi ^2) \sin ^9( \phi t )+9216 \alpha  \zeta
^7 \phi ^3 e^{7 \varphi t}\\\nonumber && \sin ^5( \phi t) \cos ^3(
\phi t) (7 \zeta  e^{ \varphi t} \sin ( \phi t)10 \varphi +8)+288
\alpha \zeta ^7 e^{7 \varphi t} (144 \varphi ^4+320 \varphi ^3-16 (9
\varphi +4)\\\nonumber && \varphi \phi ^2-1) \sin ^8( \phi t)+576
\alpha \zeta ^6 \phi ^4 e^{6 \varphi t} \sin ^4(2  \phi t) (\zeta
e^{ \varphi t}+2 \csc ( \phi t))+576 \alpha \zeta ^6 \varphi e^{6
\varphi t} \\\nonumber &&(48 \varphi ^3-16 \varphi \phi ^2-1) \sin
^7( \phi t)-96 \alpha  \zeta ^5 \varphi  (3 \varphi -8) e^{5 \varphi
t} \sin ^6(\phi t)+192 \alpha \zeta ^4 e^{4 \varphi t} (3 \varphi
^2-\phi ^2) \\\nonumber &&\sin ^5( \phi t)+2 \sin ( \phi t) (5760
\alpha \zeta ^6 \varphi \phi ^3 e^{6 \varphi t} \sin ^3(2 \phi t)+48
\alpha \zeta ^4 \phi ^2 e^{4 \varphi t} \sin ^2(2 \phi t)-\varphi
)+288 \alpha \\\nonumber &&\zeta ^5 \phi ^2 e^{5 \varphi t} \sin ^4(
\phi t) \cos ^2( \phi t) (192 \zeta ^4 e^{4 \varphi t} \sin ^4( \phi
t)+32 \zeta ^3 (31 \varphi +10) e^{3 \varphi t} \sin ^3( \phi
t)\\\nonumber && +16 \zeta ^2 e^{2 \varphi t} (\varphi (45 \varphi
+56)-7 \phi ^2) \sin ^2( \phi t)+32 \zeta e^{ \varphi t} (17 \varphi
^2-\phi ^2) \sin ( \phi t)-1)-2 \phi \cos ( \phi t)\\\nonumber &&
(-18432 \alpha \zeta ^9 (4 \varphi +1) e^{9 \varphi t} \sin ^9( \phi
t)-2304 \alpha \zeta ^8 e^{8 \varphi t} (\varphi (75 \varphi +44)-11
\phi ^2) \sin ^8(\phi t)-9216 \alpha \zeta ^7 \\\nonumber &&e^{7
\varphi t} (3 \varphi ^2 (3 \varphi +5)-(4 \varphi +1) \phi ^2) \sin
^7( \phi t)-288 \alpha \zeta ^6 e^{6  \varphi t} (192 \varphi ^3-32
\varphi \phi ^2-1)\\\nonumber && \sin ^6( \phi t)+96 \alpha \zeta ^5
(3 \varphi -4) e^{5 \varphi t} \sin ^5( \phi t)-576 \alpha \zeta ^4
\varphi e^{4 \varphi t} \sin ^4( \phi t)+1)\\\label{32} &&-3 \zeta
e^{ \varphi t} \sin ^2( \phi t)))^{\frac{1}{2}}
\end{eqnarray}
Now, the profiles of energy density and pressure under the presence
of Eq.\eqref{30}, are provided in fig.\ref{f3}. The plots indicate
that the energy density suffers a positive behavior for the assumed
values of free parameters. Similarly, the negative behavior for the
pressure term indicates that the universe is in the accelerated
expansion phase. However, the positive density proves a strong
validation for the verification of the energy conditions. Also, one
can get the positive and alternate trends of the both terms for
different time periods due to the oscillatory behavior of the
assumed Hubble parameter. We restrict our work for the positive
density and negative pressure behavior to ascertain the energy
conditions. The evolutionary profiles of the energy conditions are
provided in the figs. \ref{f4} and \ref{f5}. The $\mathcal{NEC}$
plot shows the violation with in the bouncing regime and confirms
the major verification for the universe to attain a bounce with in
the framework of $\mathcal{FLRW}$ spacetime. The violated
$\mathcal{WEC}$ and $\mathcal{SEC}$ are given in the left plots of
the figs. \ref{f3} and \ref{f4}. The violated $\mathcal{SEC}$ also
maintains the recent observations for the accelerating universe
\cite{visser1997energy}. One important energy condition i.e,
$\mathcal{TEC}$ has also been given in this recent study. The
positive profiles for the $\mathcal{DEC}$ and $\mathcal{TEC}$ are
given in the fig.\ref{f5}. The evolution of these energy conditions
is strictly dependent on the values of the free parameters used in
this study. However, one can get another configuration of these
physical factors by implementing the different free parameters. The
evolution of $\emph{EoS}$ parameter is provided in fig.\ref{f6} to
encounter the negative value i.e, $\omega(t)\approx-1$, for the
current expansion phase of the universe.
\section{Discussions}
This work involves the study of bouncing cosmology for an isotropic
configuration of fluid $\mathcal{T}_{\alpha\beta}$ and
$\mathcal{FLRW}$ metric. We comprehend this work under
$f(\mathcal{G},\mathcal{T}_{\alpha \beta} \mathcal{T}^{\alpha
\beta})$ theory of gravitation by assuming a specific model i.e,
$f(\mathcal{G},\mathcal{T}^2)=\mathcal{G}+\alpha
\mathcal{G}^2+2\lambda \mathcal{T}^2$ with $\alpha$ and $\lambda$
are constants, serving as free parameters. This is the first-ever
attempt to cover bouncing cosmology in the
$f(\mathcal{G},\mathcal{T}_{\alpha \beta} \mathcal{T}^{\alpha
\beta})$ theory. By the consideration of a specific functional form
of the Hubble parameter, we discuss the evolution of cosmographic
parameters. The assumption of a well-known equation of state
($\emph{EoS}$) parameter, $\omega(t)=-\frac{k \log (t+\epsilon
)}{t}-1$, is used as a direct implementation to represent the
dynamical behavior of energy density, matter pressure, and energy
conditions. The free parameters are restricted to the special values
provided in each graph plot and are used for  $\mathcal{H}(t)$ to
act as the bouncing solution. The viability of energy conditions is
studied with the help of a graphical approach. Following are the
concluding remarks for this present work.
\begin{itemize}
  \item The Hubble parameter $\mathcal{H}(t)$ used in this study is considered
  to have a trigonometric functional form. The evolutionary behavior of
  different cosmographic factors is described under the same form of
  $\mathcal{H}(t)$. This parameterized form of $\mathcal{H}(t)$ depends on the periodic
  values of the function $\sin ( \phi t)$ and $h(t)$. We considered
  this $h(t)$ as a nonvanishing function for the periodic values of
  $\sin ( \phi t)$. A perfect bouncing model allows the Hubble parameter to show the
contraction phase i.e, $\mathcal{H}<0$, and when the universe expands
it becomes $\mathcal{H}>0$. During this expansion and contraction
phase, there is the point in between, at where $\mathcal{H}(t)$ becomes
zero. So, in order to produce such a scenario, we have arranged the
constants ($\phi$ and $\varphi$) in the Hubble parameter
$(\mathcal{H}(t)=\zeta \sin ( \phi t) \exp(\varphi t))$ to some
specific values and notice the bounce at $t=313$. However, $t=313$
is significant in such a way that all the energy conditions
necessary for the bounce, get satisfied accordingly till
$t=313$, depending on the values of $\phi$ and $\varphi$. One can
also produce other values of $t$ for bounce by restricting other
values of $\phi$ and $\varphi$. The plot of $\mathcal{H}(t)$ is
given in
  fig.\ref{F1}. The Hubble parameter gives us the bounce at $t=313$ which is the
  future singularity in the scale factor, see fig.\ref{F1}. The mathematical
  forms of deceleration,
  jerk, and snap are evaluated with the same $\mathcal{H}(t)$. The
  deceleration parameter tends to have a negative trend i.e,
  $q(t)$ approaches $-1$, which can be seen in fig.\ref{f6}. Similarly, the trends of jerk and
  snaps are given in fig.\ref{F2} with $j(t)$ approaches to $1$ and
  $s(t)$ approaches to $0$. All these values show a deflection at
  the bouncing point, that fits in for the bouncing universe.
  \item  We ensure the configuration of the bouncing cosmology
by studying energy conditions. These energy conditions are provided
  in terms of energy density and matter pressure derived from
  the modified field equations. We assumed a specific $\emph{EoS}$ parameter in
  the form $\omega(t)=-\frac{k \log (t+\epsilon )}{t}-1$. This
  $\emph{EoS}$ parameter helped to maintain the positive and negative growth of
  energy density and matter pressure for the limited bouncing time period. The
  profiles of $\rho$ and $p$ are provided in the fig.\ref{f3}. However, the
  mathematical expression for these terms is evaluated in
  Eqs.\eqref{31} and \eqref{32}.
  \item Under the restricted values of the free parameters, $\alpha
  =0.005$, $k=0.5$, $\varphi =0.001$, $\epsilon =0.001$, $\phi =0.01$,
  $\kappa =1$ and $\lambda =-0.005$, we get the violation of the
  $\mathcal{NEC}$ and $\mathcal{SEC}$. The
  violated $\mathcal{NEC}$ derives the bouncing nature of the
  universe. However, the violated $\mathcal{SEC}$ and $\mathcal{WEC}$
  provide the phase of cosmic expansion. suitable with the
  observational data. The left plots of figs.\ref{f3} and \ref{f4}
  shows the violated $\mathcal{SEC}$ and $\mathcal{WEC}$. Similarly,
  the positive behavior of $\mathcal{DEC}$ and $\mathcal{TEC}$
  assure that the assumed model configuration is valid.
  Figure \ref{f5} represents the illustration of $\mathcal{DEC}$ and
  $\mathcal{TEC}$. Also, the evolution of $\emph{EoS}$ can be seen
  in fig.\ref{f6}, showing that $\omega(t)\rightarrow-1$. This value
  of $\omega(t)$ favors the current accelerated expansion phase of the
  universe \cite{hogan2007unseen,corasaniti2004foundations,weller2003large}.
  \item The above discussion provides that the bouncing evolution of
 the universe, studied in the framework of $f(\mathcal{G},
 \mathcal{T}^2)=\mathcal{G}+\alpha\mathcal{G}^2+2\lambda
 \mathcal{T}^2$ and agrees with the recent astronomical observations
 \cite{carloni2005cosmological,fay2007f}
 i.e, all the energy conditions are fully satisfied, a great negative
 pressure behavior had been observed and provided help to study the
 late time accelerated universe \cite{elizalde2020cosmological}. However,
  this study can be used in the future for different models of
  the scale factors and Hubble parameters.
  \item We finally conclude that the bouncing evolution of the
  universe can be studied effectively with the oscillating nature of
  the scale factor under the flat $\mathcal{FLRW}$ regime.
\end{itemize}

 \vspace{0.5cm}
\bibliography{BIBFILE}

\begin{thebibliography}{10}

\bibitem{hogan1998little}
C.~J. Hogan, {\em The little book of the big bang: A cosmic primer}.
\newblock Springer Science \& Business Media, 1998.

\bibitem{guth2001eternal}
A.~H. Guth, ``Eternal inflation,'' {\em Ann. N. Y. Acad. Sci.}, vol.~950,
  no.~1, pp.~66--82, 2001.

\bibitem{padmanabhan1988does}
T.~Padmanabhan and T.~R. Seshadri, ``Does inflation solve the horizon
  problem?,'' {\em Class. Quantum Gravity}, vol.~5, no.~1, p.~221, 1988.

\bibitem{earman1999critical}
J.~Earman and J.~Mosterin, ``A critical look at inflationary cosmology,'' {\em
  Philos. Sci.}, vol.~66, no.~1, pp.~1--49, 1999.

\bibitem{ijjas2018bouncing}
A.~Ijjas and P.~J. Steinhardt, ``Bouncing cosmology made simple,'' {\em Class.
  Quantum Grav.}, vol.~35, no.~13, p.~135004, 2018.

\bibitem{alesci2017cosmological}
E.~Alesci, G.~Botta, F.~Cianfrani, and S.~Liberati, ``Cosmological singularity
  resolution from quantum gravity: The emergent-bouncing universe,'' {\em Phys.
  Rev. D}, vol.~96, no.~4, p.~046008, 2017.

\bibitem{das2018cosmological}
P.~Das, S.~Pan, S.~Ghosh, and P.~Pal, ``Cosmological time crystal: Cyclic
  universe with a small cosmological constant in a toy model approach,'' {\em
  Phys. Rev. D}, vol.~98, no.~2, p.~024004, 2018.

\bibitem{mielczarek2010observational}
J.~Mielczarek, M.~Kamionka, A.~Kurek, and M.~Szyd{\l}owski, ``Observational
  hints on the big bounce,'' {\em J. Cosmol. Astropart. Phys.}, vol.~2010,
  no.~07, p.~004, 2010.

\bibitem{cai2013anisotropy}
Y.-F. Cai, R.~Brandenberger, and P.~Peter, ``Anisotropy in a non-singular
  bounce,'' {\em Class. Quantum Gravity}, vol.~30, no.~7, p.~075019, 2013.

\bibitem{cai2009evolution}
Y.-F. Cai and X.~Zhang, ``Evolution of metric perturbations in a model of
  nonsingular inflationary cosmology,'' {\em J. Cosmol. Astropart.}, vol.~2009,
  no.~06, p.~003, 2009.

\bibitem{roshan2016energy}
M.~Roshan and F.~Shojai, ``Energy-momentum squared gravity,'' {\em Phys. Rev.
  D}, vol.~94, no.~4, p.~044002, 2016.

\bibitem{nojiri2005modified}
S.~Nojiri and S.~D. Odintsov, ``Modified \textsc{G}auss--\textsc{B}onnet theory
  as gravitational alternative for dark energy,'' {\em Phys. Lett. B},
  vol.~631, no.~1-2, pp.~1--6, 2005.

\bibitem{astashenok2015modified}
A.~V. Astashenok, S.~D. Odintsov, and V.~K. Oikonomou, ``Modified gauss--bonnet
  gravity with the lagrange multiplier constraint as mimetic theory,'' {\em
  Class. Quantum Gravity}, vol.~32, no.~18, p.~185007, 2015.

\bibitem{sharif2016energy}
M.~Sharif and A.~Ikram, ``Energy conditions in f (\textsc{G, T}) gravity,''
  {\em Eur. Phys. J. C .}, vol.~76, no.~11, pp.~1--13, 2016.

\bibitem{bhatti2021electromagnetic}
M.~Z. Bhatti, M.~Y. Khlopov, Z.~Yousaf, and S.~Khan, ``Electromagnetic field
  and complexity of relativistic fluids in f (\textsc{G}) gravity,'' {\em Mon.
  Not. Roy. Astron. Soc.}, vol.~506, pp.~4543--4560, 2021.

\bibitem{yousaf2022f}
Z.~Yousaf, M.~Z. Bhatti, S.~Khan, and P.~K. Sahoo, ``f $(\textsc{G},
  \textsc{T}^{\alpha\beta} \textsc{T}_{\alpha\beta})$ theory and complex
  cosmological structures,'' {\em Phys. Dark Universe}, p.~101015, 2022.

\bibitem{katirci2014f}
N.~Kat{\i}rc{\i} and M.~Kavuk, ``$ f (\textsc{R},\mathcal{T}_{\mu \nu}
  \mathcal{T}^{\mu \nu}) $ gravity and cardassian-like expansion as one of its
  consequences,'' {\em Eur. Phys. J. Plus}, vol.~129, no.~8, pp.~1--12, 2014.

\bibitem{guth2007eternal}
A.~H. Guth, ``Eternal inflation and its implications,'' {\em J. Phys. A},
  vol.~40, no.~25, p.~6811, 2007.

\bibitem{steinhardt2002cosmic}
P.~J. Steinhardt and N.~Turok, ``Cosmic evolution in a cyclic universe,'' {\em
  Phys. Rev. D}, vol.~65, p.~126003, 2002.

\bibitem{ijjas2017fully}
A.~Ijjas and P.~J. Steinhardt, ``Fully stable cosmological solutions with a
  non-singular classical bounce,'' {\em Phys. Lett. B}, vol.~764, pp.~289--294,
  2017.

\bibitem{bhattacharjee2020comprehensive}
S.~Bhattacharjee and P.~K. Sahoo, ``Comprehensive analysis of a non-singular
  bounce in f (\textsc{R}, \textsc{T}) gravitation,'' {\em Phys. Dark
  Universe}, vol.~28, p.~100537, 2020.

\bibitem{bamba2014bouncing}
K.~Bamba, A.~N. Makarenko, A.~N. Myagky, and S.~D. Odintsov, ``Bouncing
  cosmology in modified gauss--bonnet gravity,'' {\em Phys. Lett. B}, vol.~732,
  pp.~349--355, 2014.

\bibitem{yousaf2022cosmic}
Z.~Yousaf, M.~Z. Bhatti, and H.~Aman, ``Cosmic bounce with $\alpha (e^{-\beta
  \textsc{G}}- 1)+ 2\lambda$ \textsc{T} model,'' {\em Phys. Scr.}, vol.~97,
  no.~5, p.~055306, 2022.

\bibitem{yousaf2022bouncing}
Z.~Yousaf, M.~Z. Bhatti, and H.~Aman, ``The bouncing cosmic behavior with
  logarithmic law $f (\textsc{G, T})$ model,'' {\em Chin. J. Phys.}, vol.~79,
  pp.~275--286, 2022.

\bibitem{visser2004jerk}
M.~Visser, ``Jerk, snap and the cosmological equation of state,'' {\em Class.
  Quantum Gravity}, vol.~21, no.~11, p.~2603, 2004.

\bibitem{gruber2014cosmographic}
C.~Gruber and O.~Luongo, ``Cosmographic analysis of the equation of state of
  the universe through pad{\'e} approximations,'' {\em Phys. Rev. D}, vol.~89,
  no.~10, p.~103506, 2014.

\bibitem{busti2015cosmography}
V.~C. Busti, A.~de~la Cruz-Dombriz, P.~K. Dunsby, and D.~Saez-Gomez, ``Is
  cosmography a useful tool for testing cosmology?,'' {\em Phys. Rev. D},
  vol.~92, no.~12, p.~123512, 2015.

\bibitem{lobo2022dynamical}
F.~S.~N. Lobo, J.~P. Mimoso, J.~Santiago, and M.~Visser, ``Dynamical analysis
  of the redshift drift in f l r w universes,'' 2022.

\bibitem{moresco20166}
Moresco {\em et~al.}, ``A 6\% measurement of the hubble parameter at z~ 0.45:
  direct evidence of the epoch of cosmic re-acceleration,'' {\em J. Cosmol.
  Astropart. Phys.}, vol.~2016, no.~05, p.~014, 2016.

\bibitem{hu2021measuring}
J.~P. Hu, F.~Y. Wang, and Z.~G. Dai, ``Measuring cosmological parameters with a
  luminosity-time correlation of gamma-ray bursts,'' {\em Mon. Not. Roy.
  Astron. Soc.}, vol.~507, no.~1, pp.~730--742, 2021.

\bibitem{wang2022standardized}
F.~Y. Wang, J.~P. Hu, G.~Q. Zhang, and Z.~G. Dai, ``Standardized long gamma-ray
  bursts as a cosmic distance indicator,'' {\em Astrophys. J.}, vol.~924,
  no.~2, p.~97, 2022.

\bibitem{krishnan2020there}
Krishnan {\em et~al.}, ``Is there an early universe solution to hubble
  tension?,'' {\em Phys. Rev. D}, vol.~102, no.~10, p.~103525, 2020.

\bibitem{font2014quasar}
Font-Riberan {\em et~al.}, ``Quasar-lyman $\alpha$ forest cross-correlation
  from boss dr11: Baryon acoustic oscillations,'' {\em J. Cosmol. Astropart.
  Phys.}, vol.~2014, no.~05, p.~027, 2014.

\bibitem{hu2022revealing}
J.-P. Hu and F.-Y. Wang, ``Revealing the late-time transition of $ h_{o}$:
  relieve the hubble crisis,'' {\em Mon. Not. Royal Astron. Soc.}, vol.~517,
  no.~1, pp.~576--581, 2022.

\bibitem{king2014high}
A.~L. King, T.~M. Davis, K.~D. Denney, M.~Vestergaard, and D.~Watson,
  ``High-redshift standard candles: predicted cosmological constraints,'' {\em
  Mon. Not. Roy. Astron. Soc.}, vol.~441, no.~4, pp.~3454--3476, 2014.

\bibitem{zhang2013cosmological}
M.-J. Zhang, C.~Ma, Z.-S. Zhang, Z.-X. Zhai, and T.-J. Zhang, ``Cosmological
  constraints on holographic dark energy models under the energy conditions,''
  {\em Phys. Rev. D}, vol.~88, no.~6, p.~063534, 2013.

\bibitem{babichev2004dark}
E.~Babichev, V.~Dokuchaev, and Y.~Eroshenko, ``Dark energy cosmology with
  generalized linear equation of state,'' {\em Class. Quantum Grav.}, vol.~22,
  no.~1, p.~143, 2004.

\bibitem{haro2015gravitational}
J.~Haro and E.~Elizalde, ``Gravitational particle production in bouncing
  cosmologies,'' {\em J. Cosmol. Astropart. Phys.}, vol.~2015, no.~10, p.~028,
  2015.

\bibitem{bacalhau2018consistent}
A.~P. Bacalhau, N.~Pinto-Neto, and S.~D.~P. Vitenti, ``Consistent scalar and
  tensor perturbation power spectra in single fluid matter bounce with dark
  energy era,'' {\em Phys. Rev. D}, vol.~97, no.~8, p.~083517, 2018.

\bibitem{melia2022friedmann}
F.~Melia, ``The
  \textsc{F}riedmann--\textsc{L}ema{\^\i}tre--\textsc{R}obertson--\textsc{W}alker
  metric,'' {\em Mod. Phys. Lett. A}, vol.~37, no.~03, p.~2250016, 2022.

\bibitem{yousaf2016causes}
Z.~Yousaf, K.~Bamba, and M.~Z. Bhatti, ``Causes of irregular energy density in
  $f(\textsc{R},\textsc{T})$ gravity,'' {\em Phys. Rev. D}, vol.~93, p.~124048,
  2016.

\bibitem{shamir2021bouncing}
M.~F. Shamir, ``Bouncing universe in f (\textsc{G, T}) gravity,'' {\em Phys.
  Dark Universe}, vol.~32, p.~100794, 2021.

\bibitem{nojiri2017modified}
S.~Nojiri, S.~D. Odintsov, and V.~K. Oikonomou, ``Modified gravity theories on
  a nutshell: inflation, bounce and late-time evolution,'' {\em Phys. Rep.},
  vol.~692, pp.~1--104, 2017.

\bibitem{elizalde2020cosmological}
E.~Elizalde, N.~Godani, and G.~C. Samanta, ``Cosmological dynamics in
  $\textsc{R}^2$ gravity with logarithmic trace term,'' {\em Phys. Dark
  Universe}, vol.~30, p.~100618, 2020.

\bibitem{sharif2015instability}
M.~Sharif and Z.~Yousaf, ``Instability of meridional axial system in
  f(\textsc{R}) gravity,'' {\em Eur. Phys. J. C}, vol.~75, p.~194, 2015.

\bibitem{bhatti2018existence}
M.~Z. Bhatti, Z.~Yousaf, and M.~Ilyas, ``Existence of wormhole solutions and
  energy conditions in f (\textsc{R,T}) gravity,'' {\em J. Astrophys. Astron.},
  vol.~39, p.~69, 2018.

\bibitem{yousaf2020definition}
Z.~Yousaf, ``Definition of complexity factor for self-gravitating systems in
  \textsc{P}alatini f(\textsc{R}) gravity,'' {\em Phys. Scr.}, vol.~95,
  p.~075307, 2020.

\bibitem{nasir2022influence}
M.~M.~M. Nasir, M.~Z. Bhatti, and Z.~Yousaf, ``Influence of \textsc{EMSG} on
  complex systems: Spherically symmetric, static case,'' {\em Int. J. Mod.
  Phys. D}, p.~10.1142/S0218271823500098.

\bibitem{shamir2021abouncing}
M.~F. Shamir, ``Bouncing cosmology in gravity with logarithmic trace term,''
  {\em Adv. Astron.}, vol.~2021, p.~8852581, 2021.

\bibitem{hu2022high}
J.~P. Hu and F.~Y. Wang, ``High-redshift cosmography: Application and
  comparison with different methods,'' {\em Astron. Astrophys.}, vol.~661,
  p.~A71, 2022.

\bibitem{hawking1973large}
S.~W. Hawking and G.~F.~R. Ellis, {\em The large scale structure of
  space-time}, vol.~1.
\newblock Cambridge university press, 1973.

\bibitem{visser1997energy}
M.~Visser, ``Energy conditions in the epoch of galaxy formation,'' {\em
  Science}, vol.~276, no.~5309, pp.~88--90, 1997.

\bibitem{nojiri2003effective}
S.~Nojiri and S.~D. Odintsov, ``Effective equation of state and energy
  conditions in phantom/tachyon inflationary cosmology perturbed by quantum
  effects,'' {\em Phys. Lett. B}, vol.~571, no.~1-2, pp.~1--10, 2003.

\bibitem{bertolami2009energy}
O.~Bertolami and M.~C. Sequeira, ``Energy conditions and stability in f
  (\textsc{R}) theories of gravity with nonminimal coupling to matter,'' {\em
  Phys. Rev. D}, vol.~79, no.~10, p.~104010, 2009.

\bibitem{balart2014regular}
L.~Balart and E.~C. Vagenas, ``Regular black hole metrics and the weak energy
  condition,'' {\em Phys. Lett. B}, vol.~730, pp.~14--17, 2014.

\bibitem{larson2011seven}
Larson {\em et~al.}, ``Seven-year wilkinson microwave anisotropy probe
  (\textsc{WMAP}*) observations: power spectra and \textsc{WMAP}-derived
  parameters,'' {\em Astrophys. J., Suppl. Ser}, vol.~192, no.~2, p.~16, 2011.

\bibitem{caldwell2002phantom}
R.~R. Caldwell, ``A phantom menace? cosmological consequences of a dark energy
  component with super-negative equation of state,'' {\em Phys. Lett. B},
  vol.~545, no.~1-2, pp.~23--29, 2002.

\bibitem{alam2004there}
U.~Alam, V.~Sahni, T.~Deep~Saini, and A.~A. Starobinsky, ``Is there supernova
  evidence for dark energy metamorphosis?,'' {\em Mon. Not. Roy. Astron. Soc.},
  vol.~354, no.~1, pp.~275--291, 2004.

\bibitem{onemli2004quantum}
V.~K. Onemli and R.~P. Woodard, ``Quantum effects can render $w<- 1$ on
  cosmological scales,'' {\em Phys. Rev. D}, vol.~70, no.~10, p.~107301, 2004.

\bibitem{yousaf2022non}
Z.~Yousaf, M.~Z. Bhatti, and S.~Khan, ``Non-static charged complex structures
  in $f(\mathcal{G},\mathcal{T}_{\alpha \beta} \mathcal{T}^{\alpha \beta})$
  gravity,'' {\em Eur. Phys. J. Plus}, vol.~137, no.~3, pp.~1--19, 2022.

\bibitem{hogan2007unseen}
J.~Hogan, ``Unseen universe: Welcome to the dark side,'' {\em Nature},
  vol.~448, no.~7151, pp.~240--246, 2007.

\bibitem{corasaniti2004foundations}
Corasaniti {\em et~al.}, ``Foundations of observing dark energy dynamics with
  the wilkinson microwave anisotropy probe,'' {\em Phys. Rev. D}, vol.~70,
  no.~8, p.~083006, 2004.

\bibitem{weller2003large}
J.~Weller and A.~M. Lewis, ``Large-scale cosmic microwave background
  anisotropies and dark energy,'' {\em Mon. Not. R. Astron. Soc}, vol.~346,
  no.~3, pp.~987--993, 2003.

\bibitem{carloni2005cosmological}
S.~Carloni, P.~K. Dunsby, S.~Capozziello, and A.~Troisi, ``Cosmological
  dynamics of $r^{n}$ gravity,'' {\em Class. Quantum Gravity}, vol.~22, no.~22,
  p.~4839, 2005.

\bibitem{fay2007f}
S.~Fay, R.~Tavakol, and S.~Tsujikawa, ``$f(\textsc{R})$ gravity theories in
  \textsc{P}alatini formalism: Cosmological dynamics and observational
  constraints,'' {\em Phys. Rev. D}, vol.~75, no.~6, p.~063509, 2007.

\end{thebibliography}
\bibliographystyle{ieeetr}

\end{document}